\documentclass[prc,superscriptaddress,showpacs,twocolumn]{revtex4}
\usepackage{graphicx,dcolumn,array,bm,amsmath}

\newcommand{\be}{\begin{equation}}
\newcommand{\ee}{\end{equation}}
\newcommand{\ba}{\begin{array}}
\newcommand{\ea}{\end{array}}
\newcommand{\bn}{\begin{eqnarray}}
\newcommand{\en}{\end{eqnarray}}
\newcommand{\bnl}{\begin{mathletters}\begin{eqnarray}}
\newcommand{\enl}{\end{eqnarray}\end{mathletters}}
\newcommand{\bml}{\begin{mathletters}}
\newcommand{\eml}{\end{mathletters}}
\newcommand{\bc}{\begin{center}}
\newcommand{\ec}{\end{center}}
\newcommand{\bi}{\begin{itemize}}
\newcommand{\ei}{\end{itemize}}
\newcommand{\bt}{\begin{tabular}}
\newcommand{\et}{\end{tabular}}
\newcommand{\bnll}[1]{\begin{subequations}\label{#1}\begin{eqnarray}}
\newcommand{\enll}{\end{eqnarray}\end{subequations}}

\newcommand{\thalf}{\tfrac{1}{2}}
\newcommand{\tthird}{\tfrac{1}{3}}
\newcommand{\tquart}{\tfrac{1}{4}}
\newcommand{\Tr}{\mbox{Tr}}
\newcommand{\rmd}{{\rm d}}

\newcommand{\intsum}{\int\hspace{-1.4em}\sum}
\newcommand{\tintsum}{{\textstyle\int\hspace{-0.9em}\sum}}
\newcommand{\hatF}{}
\newcommand{\ph}{\phantom{$-$}}
\newcommand{\pr}{\phantom{-}}

\renewcommand{\bbox}[1]{\bm{#1}}
\newcommand{\ofbboxofr}{(\bbox{r})}
\renewcommand{\ofbboxofr}{}
\newcommand{\mathsfI}{\delta}

\newcommand{\particle}{p-h}
\newcommand{\pairing} {p-p}
\newcommand{\Particle}{P-H}
\newcommand{\Pairing} {P-P}
\begin{document}

\title{Local Density Approximation for proton-neutron
pairing correlations. I. Formalism}
\author{E. Perli\'nska}
\email[]{Elzbieta.Perlinska@fuw.edu.pl}
\affiliation{Institute of Theoretical Physics,
Warsaw University,
ul. Ho\.za 69, PL-00681, Warsaw, Poland}
\affiliation{Institute of Physical Chemistry of the Polish
Academy of Sciences, ul. Kasprzaka 44/52,
PL-01224, Warsaw, Poland}
\author{S.G. Rohozi\'nski}
\email[]{Stanislaw-G.Rohozinski@fuw.edu.pl}
\affiliation{Institute of Theoretical Physics,
Warsaw University,
ul. Ho\.za 69, PL-00681, Warsaw, Poland}
\author{J. Dobaczewski}
\email[]{Jacek.Dobaczewski@fuw.edu.pl}
\affiliation{Institute of Theoretical Physics,
Warsaw University,
ul. Ho\.za 69, PL-00681, Warsaw, Poland}
\author{W. Nazarewicz}
\email[]{witek@utk.edu}
\affiliation{Institute of Theoretical Physics,
Warsaw University,
ul. Ho\.za 69, PL-00681, Warsaw, Poland}
\affiliation{Department of Physics and Astronomy,
The University of Tennessee,
Knoxville, Tennessee 37996}
\affiliation{Physics Division,
Oak Ridge National Laboratory,
P.O. Box 2008, Oak Ridge, Tennessee 37831}

\date{60th draft: October 1, 2003, today: \today}

\begin{abstract}
In the present study we generalize the self-consistent
Hartree-Fock-Bogoliubov (HFB) theory formulated in the coordinate
space to the case which incorporates an arbitrary mixing between protons and
neutrons in the particle-hole (p-h) and particle-particle (p-p or pairing)
channels. We define the HFB density matrices, discuss their
spin-isospin structure, and construct the  most general energy density functional
that is
quadratic in local densities.
The consequences of the local gauge invariance are discussed and
the particular case of the Skyrme energy density functional is studied.
By varying the total
energy with respect to the density matrices the self-consistent
one-body HFB Hamiltonian is
obtained and the structure of the resulting mean fields is shown. The
consequences of
the time-reversal symmetry, charge invariance, and proton-neutron
symmetry are summarized.
The complete list of expressions required to calculate total
energy is presented.
\end{abstract}

\maketitle

\section{Introduction}\label{sec1}

One of the main goals of nuclear theory is to build the unified
microscopic framework for heavy nuclei in which the bulk nuclear
properties, nuclear excitations, and nuclear reactions can be described on the same footing.
Microscopic theory also provides the solid foundation for
phenomenological models and coupling schemes which have been applied so
successfully to explain specific nuclear properties.
Exotic short-lived nuclei are very important in this quest. The abnormal
neutron-to-proton ratios of these nuclei
isolate and  amplify important
features, which are not clearly visible
in stable systems.

For medium-mass and heavy nuclei, a critical challenge is the quest for the universal
energy density functional, which will be able to describe properties of
finite nuclei  as well as  extended asymmetric nucleonic matter (e.g.,
as found in neutron stars).
Self-consistent  methods based on the density functional theory have  already
achieved a level of sophistication and precision which allows analyses
of experimental data for a wide range of properties and for arbitrarily
heavy nuclei. For instance,   self-consistent HF and  HFB  models
are now able to reproduce  measured  nuclear  binding energies  with an
impressive rms error of $\sim$700 keV \cite{[Gor02],[Sam02]}. However,
much work  remains to be done. Developing a universal nuclear density
functional  will require a better
understanding of the    density dependence, isospin effects, pairing,
as well as an improved treatment of many-body correlations. All those aspects
are essential for the structure  of  proton-rich nuclei with  $N$$\approx$$Z$, which
are expected to exhibit  proton-neutron (pn)
pairing \cite{[Goo79]}; it is precisely in those nuclei that the
state-of-the-art microscopic mass formula needs to be supplemented by
a phenomenological Wigner term \cite{[Gor02],[Sam02]}.

In spite of an impressive experimental progress in the
heavy  $N$$\approx$$Z$ region, it is still
unclear (i)  what the specific  fingerprints
of the pn  pairing are and (ii)  what is the interplay between the like-particle and pn
($T$=0,1) p-h and p-p  channels. Before attempting to
answer these  questions,
established theoretical models  of nuclear pairing  need to be generalized to
properly account for pn correlations.
The present work is a step
in this direction.  We propose the general HFB formalism which fully incorporates
the pn mixing on the mean-field level. The resulting density matrices
have a very rich  spin-isospin structure, which, in
the presence of static pn pairing, can produce novel
mean-fields and deformations.

The paper is organized as follows. Section~\ref{pnpairing} contains
a  brief review of the pn pairing.
Section~\ref{sec2}  discusses the density matrices (scalar, vector, and tensor), both
in the p-h and p-p channel. The discussion is based on the coordinate-space
HFB formalism
\cite{[Bul80],[Dob84],[Dob96]}, which was introduced earlier to describe
pairing correlations between like nucleons.
This method  is the tool of choice when dealing
with weakly-bound heavy nuclei \cite{[Dob97a]}.
The energy functional is constructed
in Sec.~\ref{sec5}, the associated  mean fields are  derived in
Sec.~\ref{sec6}, and  Sec.~\ref{sec7}  deals with the
resulting coordinate-space HFB equations. In the discussion
of pn pairing, the notion of self-consistent symmetries, especially those
associated with
charge invariance and  time reversal,
is crucial, and we devote Sec.~\ref{sec8} to this topic.
Finally, conclusions are contained in Sec.~\ref{sec9}.

\section{Proton-neutron pairing, a concise overview}\label{pnpairing}

A unique aspect of proton-rich
nuclei with  $N$$\approx$$Z$ is that neutrons and protons occupy the same
single-particle  orbitals. Consequently,
due to the large spatial overlaps between neutron
and proton single-particle wave functions, pn pairing is expected to be
present in those systems.

So far, the strongest evidence for enhanced pn correlations
around the $N$=$Z$ line  comes from the
measured binding energies \cite{[Jen84],[Bre90],[Cho91],[Sat97],[Sat98b],[Kan99],[Sat00],[Nee02],[Glo03]}
and the isospin structure of
the low-lying states in odd-odd nuclei
\cite{[Baz55],[Jan65],[Zel76],[Pov98],[Mar99a],[Zel98],[Vog00],[Mac00],%
[Mac00a],[Jui00],[Sat01],[Jan02],[Glo03]}.
The  pn  correlations are
also  expected to
play some role in single-beta decay \cite{[Mut89],[Hom96],[Eng99]}, double-beta decay
\cite{[Eng88],[Che93a],[Sch96a],[Sim97],[Eng97],[Civ99a],[Sar03]},
transfer reactions \cite{[Fro70],[Fro71],[Bes77],[Eva81],[And83],[Dus86],[Dob98b]}
(see, however, Ref.~\cite{[Glo03]}),
structure of low-lying collective states
 \cite{[Zaj03]}, alpha decay and alpha correlations
\cite{[Sol60],[Eic72],[Del92],[Roe98],[Dob98b],[Has00],[Kan03]},
structure of high spin states
\cite{[Pov98],[Goe72],[Goe73],[Nic78],[Mul81],[Mul82],[Kva90],[She90c],[She90d],%
[Fra94],[Sat97a],[deA97],[Dea97a],[Ter98],[Kan98a],[Fra99],%
[Fra99a],[Pet99],[Pet99a],[Fra00b],[She00],[Sat00],[Goo01],[Wys01],[Juo01],%
[Kel01],[Jen02],[oLe03],[Dob03]},
and in properties of low-density nuclear matter
\cite{[Kje68],[Pal75],[Noz85],[Elg98],[Elg98a],[Alm90],[Von91],[Sed97],[Sed00],[Gar01],[Dea03]}.

Actually, the pn pairing  is not ``the new kid on the block" but it
has a long history and is ultimately connected
to the charge invariance of the strong Hamiltonian.
(For reference, in 1932 Heisenberg introduced isotopic spin \cite{[Hei32]}
and in 1936 Wigner introduced the nuclear SU(4) supermultiplets
\cite{[Wig37]}.) An important step was the adaptation of
Racah's concept of seniority by Racah and Talmi \cite{[Rac52]},
and Flowers \cite{[Flo52]} in 1952.
In the independent  quasiparticle (BCS) picture
\cite{[Bar57]}, pairing condensate appears as a result of an
attractive interaction between quasiparticles near the Fermi surface.
The term ``nuclear
superconductivity" was first used by Pines at the 1957 Rehovot
Conference to point out that the new BCS theory might also apply
to nuclei \cite{[Lip98]}.  This was formally accomplished  in the late fifties
\cite{[Boh58],[Bel59]} and shortly afterwards
the importance  of pn pairing  was emphasized
\cite{[Bel60],[Sol60],[Flo61]} and
a number of theoretical papers dealing
with the generalization of the BCS theory to the pn pairing
case appeared \cite{[Bre63],[Flo63],[Lan64]}.

Independently, group-theoretical methods
based on the
quasi-spin formalism were developed. Many insights were gained
by simple solvable models employing symmetry-dictated interactions
\cite{[Flo64],[Ich64],[Hec65],[Par65],[Gin65],[Ell65],[Pan69],[Dus70]}.
Two families of models were used, one based on the
$j$-$j$ coupling with the symmetry SO(5) (appropriate for the
$T$=1 pairing)   and the other based on the  $L$-$S$ coupling
with the symmetry SO(8) (appropriate for the
$T$=0 and $T$=1 pairing). These models have been consecutively
developed and applied to various physically interesting
cases \cite{[Eva81],[Dus86],[Eng96],[Eng97],[Dob98b],[Pit99],[Pal01]}.
Among many  techniques used  to solve  the problem of pn pairing with
schematic interactions,
worth mentioning are the exact methods \cite{[Ric66],[And77a],[And83]}
used to describe   isovector states of a charge-independent
pairing Hamiltonian.

Properties of pn pairing (at low and high spins,
in cold and hot nuclei) have been studied within
the large-scale shell model
(diagonalization  shell-model, variational shell model,
and Monte Carlo shell model)
\cite{[Dea03],[Pet93],[Pet94],[Pet99],[Pet99a],[Dea95],[Lan96],[Duf96],%
[Lan97],[Dea97a],[Eng98],[Pov98],[Mar99a]}.
It was concluded that the
isovector pairing in the dominating $J$=0 channel
mainly acts between time-reversed states
within  the same shell. On the other hand, isoscalar  pairing can
also involve coupling  (mainly $J$=1)
between spin-orbit partners. Consequently, spin-orbit
splitting plays a crucial role in understanding the $T$=0 pairing \cite{[Pov98]}.

It is to be noted that it is by no means obvious how to extract
``pairing correlations" from the realistic shell-model calculations.
The ``pairing Hamiltonian" is an integral part of the
residual shell-model interaction.
The shell-model Hamiltonian is usually written in the
p-p representation, but it also can be transformed
to the p-h representation by means of the Pandya
transformation \cite{[Pan56]}. This means that the
high-$J$ interaction between pairs can translate into
the low-$J$
interaction in the p-h channel.  It is only in the mean-field theory that the division
into ``particle-hole" and ``particle-particle"  channels
appears naturally. One way of translating the shell-model results into
mean-field language is by means of correlators, such
as the number of $T$=0 and $T$=1 pairs in the shell-model wave function,
\cite{[Eng96],[Lan96],[Dob97e],[Pet99]}.

The extension of the Interacting Boson Models (IBM)
to the case of pn bosons had to wait until 1980, when
IBM-3 (only $T$=1 pairs
\cite{[Ell80]})  and IBM-4 [both ($T$=1, $S$=0) and ($T$=0, $S$=1) bosons
\cite{[Ell81]}] were proposed.
For  recent  applications of various algebraic models, see
Refs.~\cite{[Isa95a],[Isa97], %
[Ell96],[Eva97],[Lac97],[Sug97],[Dec98],[Gar98a],[Isa98],[Isa99],[Jui00],[Svi03]}.

An alternative strategy to the pn pairing problem  is via the mean-field approach.
Here, the major conceptional step was the proposition that
quasiparticles are mixtures not only   particles and
holes but also  protons and neutrons. The
resulting HFB quasi-particle vacuum is a superposition of wave
functions corresponding to even-even and odd-odd nuclei
with different particle numbers. Unlike in the standard
nn and pp pairing cases, the coefficients of the Bogoliubov
transformation  are, in general, complex.
Generalized Bogoliubov
transformation,  generalized gap equations, and pn pairing fields
are discussed in Refs.  \cite{[Bre63],[Gos64],[Vog64],[Lev64],[Pal64],[Gos65],[Cam65],%
[Che66],[Che66a],[Che67],[Gin68],%
[Goo68],[Bar69],[Goo70],[Wol71],%
[Goo72],[Goe72],[San75],[San76],[Mpa77],[Nic78],%
[Goo79],[Mul82],[Muh81],[Sat97a],[Ter98],[Goo98],[Fra99],[Fra99a],[Goo00],[Bes00],[Sim03]}.

The problem of the
spontaneous isospin breaking in the mean-field theory
was  realized soon after the development
of the generalized quasiparticle approach \cite{[Cam65],[Gin68],[Eic72]}.
The symmetry is broken by the
independent (separate) treatment of $T$=1
proton and neutron pairing correlations and by the
BCS quasiparticle mean field (the generalized product wave
function is not an eigenstate of isospin). Several techniques have been developed
to restore isospin. They include the Generator Coordinate Method, RPA,
Kamlah expansion, iso-cranking, and exact projection
\cite{[Cam65],[Gin68],[Eic72],[Che78],[Civ97],[Civ99a],[Bes99],[Fra99], %
[Sat01],[Sat01a],[Sat01b],[Nee02],[Glo03]}.
It is fair to say, however, that in spite of
many  attempts  to extend  the
quasiparticle approach  to incorporate the effect
of pn  correlations, no symmetry-unrestricted mean-field calculations  of pn
pairing, based on realistic effective interaction and
the isospin-conserving formalism
have been carried out.

\section{Density matrices in the isospin space}\label{sec2}

We begin with the discussion of the building blocks of the HFB theory:
one-body density matrices.
In the HFB theory,  expectation values of all observables and, in
particular, of the nuclear Hamiltonian can be expressed as
functionals of the density matrix $\hat{\rho}$ and the pairing tensor
$\hat{\kappa}$ defined as \cite{[RS80]}
\bnll{rhos}
\hat{\rho}(\bbox{r}st,\bbox{r}'s't')
&=&\langle \Psi |a_{\bbox{r}'s't'}^{+}a_{\bbox{r}st}|\Psi \rangle ,
\label{rho}\\
\hat{\kappa}(\bbox{r}st,\bbox{r}'s't')
&=& \langle \Psi |a_{\bbox{r}'s't'}a_{\bbox{r}st}|\Psi \rangle ,
\label{kappa}
\enll
where $a_{\bbox{r}st}^{+}$ and $a_{\bbox{r}st}$ create and annihilate,
respectively, nucleons at point $\bbox{r}$, spin $s$=$\pm\thalf$, and
isospin $t$=$\pm\thalf$, while $|\Psi \rangle $ is the HFB
independent-quasiparticle state.
Instead of using the antisymmetric
pairing tensor it is more convenient to introduce the {\pairing}
density matrices that can be defined in two forms, $\hat{\tilde{\rho}}$
or $\hat{\breve{\rho}}$, denoted by ``tilde'' and ``breve'', respectively:
\bnll{rhos2}
\hat{\tilde{\rho}}(\bbox{r}st,\bbox{r}'s't')
&=& -2s'\langle \Psi |a_{\bbox{r}'-s't'}a_{\bbox{r}st}|\Psi \rangle ,
\label{rhotilde}\\
\hat{\breve{\rho}}(\bbox{r}st,\bbox{r}'s't')
&=& 4s't'\langle \Psi |a_{\bbox{r}'-s'-t'}a_{\bbox{r}st}|\Psi \rangle .
\label{rhobreve}
\enll
In Ref.\ \cite{[Dob84]}, {\pairing} density matrix $\hat{\tilde{\rho}}$
was used to treat the n-n and p-p pairing correlations without the
proton-neutron mixing. It was then shown that for conserved
time-reversal symmetry $\hat{\tilde{\rho}}$ is hermitian, and leads
to {\pairing} local densities that have the structure which is analogous to
that of the {\particle} local densities. However, in the case of the
proton-neutron mixing studied here, we decided to use the {\pairing}
density matrix $\hat{\breve{\rho}}$, because it allows a more transparent
treatment of the isoscalar and isovector pairing channels. Detailed
discussion of this point will be presented in Sec.\ \ref{sec2.3} below.

With each of density matrices of Eqs.\ (\ref{rho}) and
(\ref{rhobreve}) three other matrices are associated:
\begin{itemize}
\item[--] the hermitian conjugate matrices defined as:
\bnll{hermicon}
\hat{\rho}^{+}(\bbox{r}st,\bbox{r}'s't')&=&
\hat{\rho}^{\ast} (\bbox{r}'s't',\bbox{r}st),\\
\hat{\breve{\rho}}^{+}(\bbox{r}st,\bbox{r}'s't')&=&
\hat{\breve{\rho}}^{\ast} (\bbox{r}'s't',\bbox{r}st)
\enll
\item[--] the time-reversed matrices defined as:
\bnll{timerev}
\hat{\rho}^{T}(\bbox{r}st,\bbox{r}'s't')&=&
4ss'\hat{\rho}^{\ast}(\bbox{r}\,\mbox{$-s$}t,\bbox{r}'\,\mbox{$-s$}'t'),\\
\hat{\breve{\rho}}^{T}(\bbox{r}st,\bbox{r}'s't')&=&
4ss'\hat{\breve{\rho}}^{\ast}(\bbox{r}\,\mbox{$-s$}t,\bbox{r}'\,\mbox{$-s$}'t'),
\enll
\item[--] the charge-reversed matrices defined as:
\bnll{chargerev}
\hat{\rho}^{C}(\bbox{r}st,\bbox{r}'s't')&=&
4tt'\hat{\rho}(\bbox{r}s\,\mbox{$-t$},\bbox{r}'s'\,\mbox{$-t$}'),\\
\hat{\breve{\rho}}^{C}(\bbox{r}st,\bbox{r}'s't')&=&
4tt'\hat{\breve{\rho}}(\bbox{r}s\,\mbox{$-t$},\bbox{r}'s'\,\mbox{$-t$}'),
\enll
\end{itemize}
where the asterisk stands for the complex conjugation.

Here and below we present full sets of expressions even in those
cases when they could, in principle, be replaced by verbal
descriptions. We do so in order to avoid possible confusion at the
expense of a slight increase in the length of the paper. We think
that such an approach is highly beneficial to the reader, because in
many cases small but significant differences appear in expressions
that otherwise could have seemed analogous to one another.

The
charge-reversal operation $C$ defined in Eq.\ (\ref{chargerev}) exchanges
the neutron and proton charges, or equivalently, flips their isospin
projections. Note that the time reversal is antilinear while the
charge reversal is a linear operation, and that they commute with one
another. Symmetries of the density matrices can be conveniently
expressed in terms of just the hermitian conjugation, and time and charge
reversals. Namely, it follows from definitions (\ref{rho})
and (\ref{rhobreve}) that
\bnll{hc}
\label{hc1}
\hat{\rho}^{+} &=& \hat{\rho},
\\ \label{hc2}
\hat{\breve{\rho}}^{+} &=& -\hat{\breve{\rho}}^{TC},
\enll
where the superscript $TC$ denotes superposition of
the time (\ref{timerev}) and charge (\ref{chargerev}) reversals.

For $|\Psi \rangle $ being an independent-quasiparticle state the
density matrices  fulfill the following kinematical conditions
\bnll{proj}
\label{proj1}
\hat{\rho}\bullet \hat{\breve{\rho}}
&=&\hat{\breve{\rho}}\bullet \hat{\rho}^{TC} , \\
\hat{\rho}
&=&\hat{\rho}\bullet \hat{\rho}
+\hat{\breve{\rho}}\bullet \hat{\breve{\rho}}^{+},
\label{proj2}
\enll
where $\bullet $ stands for integration over spatial coordinates and
summation over spin and isospin indices, denoted by $\tintsum{\rm d}x$,
e.g.:
\begin{widetext}
\begin{equation}
(\hat{\rho} \bullet \hat{\breve{\rho}})(\bbox{r}_{1}s_1t_1,\bbox{r}_2s_2t_2)
=(\hat{\rho} \bullet \hat{\breve{\rho}})(x_1,x_2)=\intsum {\rm d}x\, \hat{\rho} (x_1,x)\hat{\breve{\rho}}(x,x_2)
=\int {\rm d}^3\bbox{r}\sum_{st}\hat{\rho} (\bbox{r}_1s_1t_1,
\bbox{r}st)\hat{\breve{\rho}}(\bbox{r}st,\bbox{r}_2s_2t_2),
\end{equation}
\end{widetext}
where we also abbreviated the space-spin-isospin variables by
$x$$\equiv$$\{\bbox{r}st\}$.
Equations (\ref{proj}) secure
the projectivity of the generalized density matrix:
\begin{equation}
\hat{\breve{\mathcal R}}=\hat{\mathcal W}\hat{\mathcal R}\hat{\mathcal W}^+
=\left(\begin{array}{cc}\hat{\rho}&\hat{\breve{\rho}}\\
\hat{\breve{\rho}}^{+} & \hat{1}-\hat{\rho}^{TC}\end{array}\right),
\label{genden}
\end{equation}
where $\hat{1} := \delta(x-x') := \delta
(\bbox{r}-\bbox{r}')\delta_{ss'}\delta_{tt'}$ and the unitary matrix $\hat{\mathcal W}$,
\begin{equation}
\hat{\mathcal W}
=\left(\begin{array}{cc}\hat{1}&0\\
0 & -\hat{\bbox{\sigma}}_y\hat{\tau}_2\end{array}\right),
\label{genuni}
\end{equation}
transforms the standard generalized density matrix $\hat{\mathcal R}$ (cf.\ Ref.\ \cite{[RS80]})
to the ``breve'' representation.

When the pairing correlations of only like nucleons are taken into
account, none but the diagonal (off-diagonal) matrix elements of
density matrix $\hat{\rho}$ ($\hat{\breve{\rho}}$) in isospin indices
are considered. However, in a general case of pairing correlations
between both, like and unlike nucleons, the remaining matrix elements
become relevant as well. Therefore, in the following subsections we specify
the spin-isospin structure of the {\particle} and {\pairing} density
matrices explicitly.

\subsection{Non-local densities}\label{sec2.1}

The density matrices in the spin and isospin spaces can be expressed
as linear combinations of the unity and
Pauli matrices. To write the corresponding formulae
the following notation is assumed. Vectors and vector operators in
the physical three-dimensional space are denoted with boldface
symbols, e.g., $\bbox{r}$ or $\bbox{\nabla}$, and the second rank
tensors -- with sans serif symbols, e.g., $\mathsf{J}$. Scalar
products of three-dimensional space vectors are, as usual, denoted
with the central dot: $\bbox{r}$$\cdot$$\bbox{\nabla}$. The
components of vectors and tensors are labelled with indices $a,b,c$
and the names of axes are $x$, $y$, and $z$, e.g.,
$\bbox{r}$=$(\bbox{r}_x,\bbox{r}_y,\bbox{r}_z)$. In order to make a clear distinction,
vectors in isospace are denoted with arrows and scalar products of
them --- with the circle: $\vec{v}\circ\vec{w}$. The components of
isovectors are labelled with indices $i,k$, and the names of iso-axes
are 1, 2, and 3, e.g., $\vec{v}$=$(v_1,v_2,v_3)$. Finally, isoscalars
are marked with subscript ``0'', and we often combine formulae for
isoscalars and isovectors by letting the indices run through all
the four values, e.g., $k$=0,1,2,3.

With this convention the density matrices have the following form
\begin{widetext}
\bnll{izo}
\hat{\rho}(\bbox{r}st,\bbox{r}'s't')
&=&\tquart\rho_0(\bbox{r},\bbox{r}')\delta_{ss'}\delta_{tt'}
+ \tquart\delta_{ss'}\vec{\rho}(\bbox{r},
\bbox{r}')\circ \hat{\vec{\tau}}_{tt'}
+ \tquart\bbox{s}_0(\bbox{r},\bbox{r}')\cdot \hat{\bbox{\sigma}}_{ss'}\delta_{tt'}
+ \tquart\vec{\bbox{s}}(\bbox{r},\bbox{r}')\cdot \hat{\bbox{\sigma}}_{ss'}\circ \hat{\vec{\tau}}_{tt'},
\label{izo01} \\
\hat{\breve{\rho}}(\bbox{r}st,\bbox{r}'s't')
&=&\tquart\breve{\rho}_0(\bbox{r},\bbox{r}')\delta_{ss'}\delta_{tt'}
+ \tquart\delta_{ss'}\vec{\breve{\rho}}(\bbox{r},\bbox{r}')\circ \hat{\vec{\tau}}_{tt'}
+ \tquart\breve{\bbox{s}}_0(\bbox{r},\bbox{r}')\cdot \hat{\bbox{\sigma}}_{ss'}\delta_{tt'}
+ \tquart\vec{\breve{\bbox{s}}}(\bbox{r},\bbox{r}')\cdot\hat{\bbox{\sigma}}_{ss'}\circ \hat{\vec{\tau}}_{tt'},
\label{izo02}
\enll
\end{widetext}
where $\hat{\vec{\tau}}_{tt'}=(\hat{\tau}^1_{tt'},\hat{\tau}^2_{tt'},\hat{\tau}^3_{tt'})$ and $\hat{\bbox{\sigma}}_{ss'}
=(\hat{\bbox{\sigma}}^x_{ss'},\hat{\bbox{\sigma}}^y_{ss'},\hat{\bbox{\sigma}}^z_{ss'})$ are
the isospin and spin Pauli matrices, respectively, which are accompanied by the corresponding
unity matrices, $\hat{\tau}^0_{tt'}=\delta _{tt'}$ and $\hat{\sigma}^u_{ss'}=\delta _{ss'}$. The density
matrices defined in Eqs.\ (\ref{rho}) and (\ref{rhobreve}) are now
expressed by several functions of the pair of position vectors
$\bbox{r}$ and $\bbox{r}'$. To avoid confusion, the functions
appearing on the right-hand sides of Eqs.\ (\ref{izo}) will be
called the (non-local) density functions or, simply, densities, unlike
the density matrices of  Eqs.\ (\ref{rho}) and (\ref{rhobreve})
appearing on  the left-hand sides.

The densities are traces in spin and isospin indices of the following
combinations of the density and the Pauli matrices:

\vspace*{1ex}\noindent$\bullet$
scalar densities:
\begin{itemize}
\item[--] {\particle} isoscalar and isovector densities:
\bnll{g1-2}
\rho_0(\bbox{r},\bbox{r}') &=&\sum_{st}\hat{\rho}
(\bbox{r}st,\bbox{r}'st),  \label{g1} \\
\vec{\rho}(\bbox{r},\bbox{r}')
&=&\sum_{stt'}\hat{\rho} (\bbox{r}st,%
\bbox{r}'st')\hat{\vec{\tau}}_{t't},\label{g2}
\enll
\item[--] {\pairing} isoscalar and isovector densities:
\bnll{gp1-2}
\breve{\rho}_0(\bbox{r},\bbox{r}')
&=&\sum_{st}\hat{\breve{\rho}}(\bbox{r}st,\bbox{r}'%
st), \label{gp1} \\
\vec{\breve{\rho}}(\bbox{r},\bbox{r}')
&=&\sum_{stt'}\hat{\breve{\rho}}(\bbox{r}st,%
\bbox{r}'st')\hat{\vec{\tau}}_{t't}, \label{gp2}
\enll
\end{itemize}

\noindent$\bullet$
vector densities:
\begin{itemize}
\item[--] {\particle} spin isoscalar and isovector densities:
\bnll{s1-2}
\bbox{s}_0(\bbox{r},\bbox{r}')
&=&\sum_{ss't}\hat{\rho}(\bbox{r}st,%
\bbox{r}'s't)\hat{\bbox{\sigma}}_{s's}, \label{s1} \\
\vec{\bbox{s}}(\bbox{r},\bbox{r}')
&=&\sum_{ss'tt'}\hat{\rho}(\bbox{r}st,%
\bbox{r}'s't')\hat{\bbox{\sigma}}_{s's}\hat{\vec{\tau}}_{t't}, \label{s2}
\enll
\item[--] {\pairing} spin isoscalar and isovector densities:
\bnll{sp1-2}
\breve{\bbox{s}}_0(\bbox{r},\bbox{r}')
&=&\sum_{ss't}\hat{\breve{\rho}}(\bbox{r}st,%
\bbox{r}'s't)\hat{\bbox{\sigma}}_{s's}, \label{sp1} \\
\vec{\breve{\bbox{s}}}(\bbox{r},\bbox{r}')%
&=&\sum_{ss'tt'}\hat{\breve{\rho}}(\bbox{r}%
st,\bbox{r}'s't')%
\hat{\bbox{\sigma}}_{s's}\hat{\vec{\tau}}_{t't}.  \label{sp2}
\enll
\end{itemize}

Since the {\particle} density matrix and the Pauli matrices are both
hermitian, all the {\particle} densities are hermitian too,
\bnll{hermicon2}
\rho_0(\bbox{r},\bbox{r}')
&=&\rho^{*}_0(\bbox{r}',\bbox{r}), \\
\vec{\rho}(\bbox{r},\bbox{r}')
&=&\vec{\rho}^{\,*}(\bbox{r}',\bbox{r}), \\
\bbox{s}_0(\bbox{r},\bbox{r}')
&=&\bbox{s}^{*}_0(\bbox{r}',\bbox{r}), \\
\vec{\bbox{s}}(\bbox{r},\bbox{r}')
&=&\vec{\bbox{s}}^{\,*}(\bbox{r}',\bbox{r}),
\enll
and hence, their real parts are symmetric, while the imaginary parts
are antisymmetric, with respect to transposition of spatial arguments
$\bbox{r}$ and $\bbox{r}'$.

On the other hand, the unity matrices $\hat{\sigma}^u_{ss'}=\delta_{ss'}$ and
$\hat{\tau}^0_{tt'}=\delta_{tt'}$ (scalar and isoscalar) are
$TC$-symmetric, while the vector and isovector Pauli matrices are
$TC$-antisymmetric, i.e.,
\bnll{TC}
\hat{\bbox{\sigma}}_{ss'} &=& - 4ss'\hat{\bbox{\sigma}}^*_{-s\,-s'},  \\
\hat   {\vec{\tau}}_{tt'} &=& - 4tt'\hat   {\vec{\tau}}^*_{-t\,-t'}.
\enll
We should stress here again that operation $TC$ is antilinear, and
therefore, complex conjugation appears in all right-hand-sides of
Eqs.\ (\ref{TC}), although only the Pauli matrices $\sigma_y$ and
$\tau_2$ are imaginary.

Since the {\pairing} density matrix transforms under $TC$ as in Eq.\
(\ref{hc2}), the {\pairing} densities are either symmetric
(scalar-isovector and vector-isoscalar) or antisymmetric
(scalar-isoscalar and vector-isovector) under the transposition of
their arguments, namely:
\bnll{symepai}
\label{symepai-a}
\breve{\rho}_0        (\bbox{r},\bbox{r}')
&=& - \breve{\rho}_0                (\bbox{r}',\bbox{r}), \\
\label{symepai-b}
\vec{\breve{\rho}}    (\bbox{r},\bbox{r}')
&=& \phantom{-}  \vec{\breve{\rho}} (\bbox{r}',\bbox{r}), \\
\label{symepai-c}
\breve{\bbox{s}}_0    (\bbox{r},\bbox{r}')
&=&  \phantom{-} \breve{\bbox{s}}_0 (\bbox{r}',\bbox{r}), \\
\label{symepai-d}
\vec{\breve{\bbox{s}}}(\bbox{r},\bbox{r}')
&=& - \vec{\breve{\bbox{s}}}        (\bbox{r}',\bbox{r}) .
\enll
Equations (\ref{hermicon2}) and (\ref{symepai}) are fulfilled
independently of any other symmetries conserved by the system;
they result from general properties (\ref{hc}) of density matrices
$\hat{\rho}$ and $\hat{\breve{\rho}}$.

\subsection{Local densities}\label{sec2.2}

In the HFB theory with the zero-range
Skyrme interaction \cite{[Sky59],[Vau72]},
or in the local density approximation (LDA) (cf.\ Refs.\
\cite{[Neg72],[RS80]}), the energy functional depends only on local
densities, and on local densities built from derivatives up to the
second order. These local densities are obtained by setting
$\bbox{r}'$=$\bbox{r}$ in Eqs.\ (\ref{g1-2})--(\ref{sp1-2}) {\em
after} the derivatives are performed. They will be denoted by having
one spatial argument to distinguish them from the non-local densities
that have two. Moreover, for local densities the spatial argument
will often be omitted in order to lighten the notation.

Following the standard definitions \cite{[Flo75],[Eng75]},
a number of local densities are introduced:

\vspace{1ex}\noindent$\bullet$
scalar densities:
\begin{itemize}
\item[--]  particle and pairing densities:
\bnll{scalar-p}
\label{scalar-p-ph}
{\rho}_k(\bbox{r})
&=&           {\rho}_k(\bbox{r},\bbox{r}')_{\bbox{r}=\bbox{r}'},\\
\label{scalar-p-pp}
\vec{\breve       {\rho}} (\bbox{r})
&=&\vec{\breve{\rho}} (\bbox{r},\bbox{r}')_{\bbox{r}=\bbox{r}'},
\enll
\item[--]  {\particle} and {\pairing} kinetic densities:
\bnll{scalar-k}
\label{scalar-k-ph}
{\tau}_k(\bbox{r})
&=&\big[        (\bbox{\nabla}\cdot\bbox{\nabla}')
{\rho}_k(\bbox{r},\bbox{r}')\big]_{\bbox{r}=\bbox{r}'}, \\
\label{scalar-k-pp}
\vec{\breve {\tau}} (\bbox{r})
&=&\big[        (\bbox{\nabla}\cdot \bbox{\nabla}')
\vec{\breve {\rho}} (\bbox{r},\bbox{r}')\big]_{\bbox{r}=\bbox{r}'},
\enll
\end{itemize}

\noindent$\bullet$
vector densities:
\begin{itemize}
\item[--]   {\particle} and {\pairing} spin (pseudovector) densities:
\bnll{vector-s}
{\bbox{s}}_k(\bbox{r})
&=&      {\bbox{s}}_k(\bbox{r},\bbox{r}')_{\bbox{r}=\bbox{r}'}, \\
\breve{\bbox{s}}_0(\bbox{r})
&=&\breve{\bbox{s}}_0(\bbox{r},\bbox{r}')_{\bbox{r}=\bbox{r}'},
\enll
\item[--]   {\particle} and {\pairing} spin-kinetic (pseudovector) densities:
\bnll{vector-T}
\label{vector-T-ph}
{\bbox{T}}_k(\bbox{r})
&=&\big[             (\bbox{\nabla}\cdot\bbox{\nabla}')
{\bbox{s}}_k(\bbox{r},\bbox{r}')\big]_{\bbox{r}=\bbox{r}'}, \\
\label{vector-T-pp}
\breve{\bbox{T}}_0(\bbox{r})
&=&\big[             (\bbox{\nabla}\cdot\bbox{\nabla}')
\breve{\bbox{s}}_0(\bbox{r},\bbox{r}')\big]_{\bbox{r}=\bbox{r}'},
\enll
\item[--]  {\particle} and {\pairing} current (vector) densities:
\bnll{vector-j}
\label{vector-j-ph}
{\bbox{j}}_k(\bbox{r})
&=&\tfrac{1}{2i}\big[ (\bbox{\nabla} - \bbox{\nabla}')
{\rho}_k(\bbox{r},\bbox{r}')\big]_{\bbox{r}=\bbox{r}'}, \\
\label{vector-j-pp}
\breve{\bbox{j}}_0(\bbox{r})
&=&\tfrac{1}{2i}\big[ (\bbox{\nabla} - \bbox{\nabla}')
\breve    {\rho}_0(\bbox{r},\bbox{r}')\big]_{\bbox{r}=\bbox{r}'},
\enll
\item[--]   {\particle} and {\pairing} tensor-kinetic (pseudovector) densities:
\bnll{vector-F}
\label{vector-F-ph}
{\bbox{F}}_k(\bbox{r})
&\!=\!&\thalf\big[       (\bbox{\nabla} \!\otimes\!\bbox{\nabla}'
\!+\!\bbox{\nabla}'\!\otimes\!\bbox{\nabla})\!\cdot\!
{\bbox{s}}_k(\bbox{r},\bbox{r}')\big]_{\bbox{r}=\bbox{r}'},\! \\
\label{vector-F-pp}
\breve{\bbox{F}}_0(\bbox{r})
&\!=\!&\thalf\big[       (\bbox{\nabla} \!\otimes\!\bbox{\nabla}'
\!+\!\bbox{\nabla}'\!\otimes\!\bbox{\nabla})\!\cdot\!
\breve{\bbox{s}}_0(\bbox{r},\bbox{r}')\big]_{\bbox{r}=\bbox{r}'},\!
\enll
\end{itemize}

\noindent$\bullet$
tensor densities:
\begin{itemize}
\item[--] {\particle} and {\pairing} spin-current (pseudotensor) densities:
\bnll{tensor-J}
\label{tensor-J-ph}
{{\mathsf J}}_k(\bbox{r})
&=&\tfrac{1}{2i}\big[ (\bbox{\nabla} - \bbox{\nabla}')\otimes
{\bbox{s}}_k(\bbox{r},\bbox{r}')\big]_{\bbox{r}=\bbox{r}'}, \\
\label{tensor-J-pp}
\vec{\breve{\mathsf J}}  (\bbox{r})
&=&\tfrac{1}{2i}\big[(\bbox{\nabla} - \bbox{\nabla}')\otimes
\vec {\breve{\bbox{s}}}  (\bbox{r},\bbox{r}')\big]_{\bbox{r}=\bbox{r}'},
\enll
\end{itemize}
where $k$=0,1,2,3, and $\otimes$ stands for the tensor product of vectors in the
physical space, e.g.,
$(\bbox{v}$$\otimes\,$$\bbox{w})_{ab}$$\,\equiv\,$$\bbox{v}_{a}\bbox{w}_{b}$ and
$[(\bbox{v}$$\otimes\,$$\bbox{w})$$\cdot$$\bbox{z}]_{a}$$\,\equiv\,$$\bbox{v}_{a}(\bbox{w}$$\cdot$$\bbox{z})$.\
Note that for particle, pairing, kinetic, spin, spin-kinetic, and
tensor-kinetic densities only the symmetric non-local densities contribute,
while for the current and spin-current densities only antisymmetric
ones contribute. It is then clear that for each {\particle} density
there exist both isoscalar and isovector component, while for the
{\pairing} densities, the isovector component exists only for the
pairing, kinetic, and spin-current densities, while the isoscalar one
exists only for spin, spin-kinetic, tensor-kinetic, and current
densities.

We note here in passing that the complete list of all local densities
(up to the derivatives of the second order) also includes the kinetic
and spin-kinetic densities in which the two derivatives are coupled
to a tensor, i.e., $\bbox{\nabla}$$\otimes$$\bbox{\nabla}'$. The
resulting local densities are usually disregarded, because they do not
have counterparts to form useful terms in the local energy density.
There is one set of exceptions, which has been overlooked in the
systematic construction presented in Ref.\ \cite{[Dob96b]}, and
appears in the averaging of a zero-range tensor force \cite{[Flo75]},
namely, the set of the tensor-kinetic local densities
(\ref{vector-F}). In Sec.\ \ref{sec5} we define terms in
the energy density that depend on the tensor-kinetic densities.

All tensor densities (\ref{tensor-J}) can be decomposed into trace,
antisymmetric, and symmetric parts, giving the standard pseudoscalar,
vector, and pseudotensor components that we show here to fix the
notation:
\bnll{tensor-s}
{J}_k(\bbox{r})&=&\sum_{a=x,y,z}           {\mathsf J}_{kaa}(\bbox{r}), \\
\vec{\breve{J}} (\bbox{r})&=&\sum_{a=x,y,z}\vec{\breve{\mathsf J}}_{aa}(\bbox{r}),
\enll
\bnll{tensor-v}
{\bbox{J}}_{ka}(\bbox{r})
&=&\sum_{b,c=x,y,z}\epsilon_{abc}           {\mathsf J}_{kbc}(\bbox{r}), \\
\vec{\breve{\bbox{J}}}_{a}(\bbox{r})
&=&\sum_{b,c=x,y,z}\epsilon_{abc}\vec{\breve{\mathsf J}}_{bc}(\bbox{r}),
\enll
\bnll{tensor-t}
\underline            {{\mathsf J}}_{kab} (\bbox{r})
&=&\thalf            {{\mathsf J}}_{kab} (\bbox{r})
+ \thalf            {{\mathsf J}}_{kba} (\bbox{r})
- \tthird                    {J}_k      (\bbox{r})\delta_{ab}, \\
\underline{\vec{\breve{{\mathsf J}}}}_{ab}(\bbox{r})
&=&\thalf \vec{\breve{{\mathsf J}}}_{ab} (\bbox{r})
+ \thalf \vec{\breve{{\mathsf J}}}_{ba} (\bbox{r})
- \tthird\vec{\breve{         J}}       (\bbox{r})\delta_{ab},
\enll
where $k$=0,1,2,3.

It follows from
Eqs.\ (\ref{hermicon2}) and (\ref{symepai}) that
the {\particle} densities are all real whereas the
{\pairing} densities are in general complex and thus the
complex-conjugate densities are relevant. The {\pairing} densities
become real or imaginary only when the time-reversal symmetry is
conserved, see Sec.\ \ref{sec8}.

Instead of the isoscalar and the third component of isovector
{\particle} density
one can always use the neutron and the proton one, e.g.,
\bnll{npden}
\rho_n(\bbox{r})&=&\thalf\left(\rho_0(\bbox{r})+\rho_3(\bbox{r})\right),
\label{nden} \\
\rho_p(\bbox{r})&=&\thalf\left(\rho_0(\bbox{r})-\rho_3(\bbox{r})\right),
\label{pden}
\enll
and just the same for all other {\particle} densities.
Similarly, instead of the $k$=1,2 isovector {\pairing} densities
one can use the neutron and proton pairing density, i.e.,
\bnll{npdenp}
\breve{\rho}_n(\bbox{r})
&=&\thalf\left(\breve{\rho}_1(\bbox{r})+i\breve{\rho}_2(\bbox{r})\right),
\label{ndenp}\\
\breve{\rho}_p(\bbox{r})
&=&\thalf\left(\breve{\rho}_1(\bbox{r})-i\breve{\rho}_2(\bbox{r})\right),
\label{pdenp}
\enll
and just the same for all other {\pairing} densities.

\subsection{The {\boldmath$TC$\unboldmath} symmetry}\label{sec2.3}

When constructing the energy density functional (Sec.\ \ref{sec5})
from the local densities (\ref{scalar-p})--(\ref{tensor-J}) one should
ensure that it is invariant with respect to: 1$^\circ$ spatial
rotations, 2$^\circ$ isospin rotations, 3$^\circ$ space inversion,
and 4$^\circ$ time reversal. All the local densities of Sec.\
\ref{sec2.2} have definite transformation properties with respect to
the first three of those, 1$^\circ$--3$^\circ$, so one can easily
construct the corresponding invariants by multiplying densities of
the same type by one another. For example, a product of any
pseudovector-isoscalar density with itself, or with any other
pseudovector-isoscalar density, is an invariant.

The time-reversal symmetry cannot be immediately treated on the same
footing, because the time-reversal and the isospin rotations do not
commute. However, as noted in Ref.\ \cite{[Boh69]}, for problems
involving the isospin symmetry it is more convenient to use the $TC$
symmetry instead of the time-reversal. Indeed, since the
charge-reversal $C$ is equivalent to a rotations by the angle $\pi$
about the iso-axis $k$=2, for conserved isospin the conservation of
$TC$ is equivalent to conservation of $T$ alone. Therefore, in order
to construct the energy density which is also time-reversal
invariant, we should classify the local densities according to the
$TC$ symmetry and then multiply by one another only densities with
the same $TC$ transformation properties.

To this end, we split the {\particle} and {\pairing} density matrices
into parts that are symmetric and antisymmetric with respect to the
$TC$ reversal, i.e., explicitly,
\bnll{TCpm}
\hat{      {\rho}}_\pm(x,x')
&=& \thalf\Big(\hat{      {\rho}}(x,x') \pm
16ss'tt'\hat{      {\rho}}^*(\bar{x},\bar{x}')\Big) , \\
\hat{\breve{\rho}}_\pm(x,x')
&=& \thalf\Big(\hat{\breve{\rho}}(x,x') \pm
16ss'tt'\hat{\breve{\rho}}^*(\bar{x},\bar{x}')\Big) ,
\enll
where we used a short-hand notation of
$\bar{x}$$\equiv$$\{\bbox{r},$$-s,$$-t\}$. In conjunction with
the $TC$ transformation properties of the Pauli matrices (\ref{TC}),
one than immediately obtains that the corresponding non-local densities
of Sec.\ \ref{sec2.1} are either real or imaginary, i.e.,

\bnll{TCcon1}
\rho_{0\pm}                      (\bbox{r},\bbox{r}')
&=&\pm\rho^{*}_{0\pm}                       (\bbox{r},\bbox{r}'), \\
\vec{\rho}_\pm                   (\bbox{r},\bbox{r}')
&=&\mp\vec{\rho}^{\,*}_\pm                  (\bbox{r},\bbox{r}'), \\
\bbox{s}_{0\pm}                  (\bbox{r},\bbox{r}')
&=&\mp\bbox{s}^{*}_{0\pm}                   (\bbox{r},\bbox{r}'), \\
\vec{\bbox{s}}_\pm               (\bbox{r},\bbox{r}')
&=&\pm\vec{\bbox{s}}^{\,*}_\pm              (\bbox{r},\bbox{r}'),
\enll
and
\bnll{TCcon2}
\breve{\rho}_{0\pm}              (\bbox{r},\bbox{r}')
&=&\pm\breve{\rho}^{*}_{0\pm}               (\bbox{r},\bbox{r}'), \\
\vec{\breve{\rho}}_\pm           (\bbox{r},\bbox{r}')
&=&\mp           \vec{\breve{\rho}}^{*}_\pm (\bbox{r},\bbox{r}'), \\
\breve{\bbox{s}}_{0\pm}          (\bbox{r},\bbox{r}')
&=&\mp           \breve{\bbox{s}}^{*}_{0\pm}(\bbox{r},\bbox{r}'), \\
\vec{\breve{\bbox{s}}}_\pm       (\bbox{r},\bbox{r}')
&=&\pm\vec{\breve{\bbox{s}}}^{*}_\pm        (\bbox{r},\bbox{r}') .
\enll

This result shows that real and imaginary parts of non-local
densities (\ref{g1-2})--(\ref{sp1-2}) have opposite $TC$
transformation properties. From Eqs.\ (\ref{TCcon1}) one then obtains
classification of local {\particle} densities, namely, the isoscalar
densities $\rho_0(\bbox{r})$, $\tau_0(\bbox{r})$, and ${\mathsf
J}_0(\bbox{r})$ are $TC$ symmetric, and $\bbox{s}_0(\bbox{r})$,
$\bbox{T}_0(\bbox{r})$, $\bbox{F}_0(\bbox{r})$, and
$\bbox{j}_0(\bbox{r})$ are $TC$ antisymmetric, while the isovector
densities $\vec{\rho}(\bbox{r})$, $\vec{\tau}(\bbox{r})$, and
$\vec{{\mathsf J}}(\bbox{r})$ are $TC$ antisymmetric, and
$\vec{\bbox{s}}(\bbox{r})$, $\vec{\bbox{T}}(\bbox{r})$,
$\vec{\bbox{F}}(\bbox{r})$, and $\vec{\bbox{j}}(\bbox{r})$ are $TC$
symmetric.

The rules of constructing the {\particle} energy density are thus
identical to those valid in the case of no proton-neutron mixing
\cite{[Eng75]}. On the other hand, from Eqs.\ (\ref{TCcon2}) one
obtains classification of local {\pairing} densities, namely, real
parts of all {\pairing} densities are $TC$ antisymmetric and imaginary
parts are $TC$ symmetric. The {\pairing} energy density must
therefore be built by multiplying real parts of different densities
with one another, and separately imaginary parts also with one
another. These rules are at the base of the energy density functional
constructed in Sec.\ \ref{sec5}.

\section{The Energy Density Functional}\label{sec5}

In the HFB theory the expectation value of Hamiltonian in state
$|\Psi\rangle$ is a functional of the density matrices, and reads
\begin{widetext}
\begin{equation}
E_{\text{HFB}}
=  \langle \Psi |\hatF{H}|\Psi \rangle
=  \overline{H}[\hat{\rho},\hat{\breve{\rho}},\hat{\breve{\rho}}^+]
=  \Tr\left(\hat{T}\bullet \hat{\rho}\right)+\thalf\Tr\left(\hat{\Gamma}\bullet \hat{\rho}+%
\hat{\breve{\Gamma}}\bullet \hat{\breve{\rho}}^{+}\right),
\label{energy}
\end{equation}
where $\Tr$ denotes integration over spatial coordinates and
summation over spin and isospin indices.
Nuclear many-body Hamiltonian $\hatF{H}$,
\begin{equation}\label{hamiltonian}
\hatF{H} = \intsum\rmd{x'}\rmd{x}\,\hat{T}(x',x)a^+_{x'}a_x
+ \tquart\intsum\rmd{x'_1}\rmd{x'_2}\rmd{x_1}\rmd{x_2}\,
\hat{V}(x'_1x'_2,x_1x_2)a^+_{x'_1}a^+_{x'_2}a_{x_2} a_{x_1},\!
\end{equation}
\end{widetext}
is composed of one-body kinetic energy $\hatF{T}$ and two-body
interaction $\hatF{V}$, being expressed in (\ref{hamiltonian}) by
matrix $\hat{T}(x',x)$ and the
antisymmetrized matrix elements $\hat{V}(x'_1x'_2,x_1x_2)$, respectively.
Matrices $\hat{\Gamma}$ and $\hat{\breve{\Gamma}}$ are the
single-particle (p-h) and pairing (p-p) self-consistent potentials,
respectively,
\bn
\hspace*{-2.0em}
\hat       {\Gamma}(x'_1,x_1)
\!\!&=&\! \intsum \rmd{x'_2}\rmd{x_2}\,
\hat{V}_{\text{p-h}}(x'_1x'_2,x_1x_2)
\hat{\rho}(x_2,x'_2), \label{partpot} \\
\hspace*{-2.0em}
\hat{\breve{\Gamma}}(x'_1,x'_2)
\!\!&=&\! \intsum \rmd{x_1}\rmd{x_2}
F_2\hat{V}_{\text{p-p}}(x'_1\bar{x}'_2,x_1\bar{x}_2)
\hat{\breve{\rho}}(x_1,x_2),
\label{pairpot}
\en
where $F_2$=$8s'_2 s_2 t'_2 t_2$ and
$\bar{x}$$\equiv$$\{\bbox{r},$$-s,$$-t\}$. In Eqs.\
(\ref{partpot}) and (\ref{pairpot}) we have indicated that the
{\particle} and {\pairing} potentials can be determined by {\em
different} two-body interactions, $\hatF{V}_{\text{p-h}}$ and
$\hatF{V}_{\text{p-p}}$, called effective interactions in the
p-h and the p-p channel, respectively. This
places further derivations in the framework of the energy-density
formalism that is not based on a definite Hamiltonian
(\ref{hamiltonian}). Moreover, effective interactions,
$\hatF{V}_{\text{p-h}}$ and $\hatF{V}_{\text{p-p}}$, are supposed to
be, in general, density-dependent.

In the case of the Skyrme  effective interaction, as well as
in the framework of the LDA, the energy functional of Eq.\
(\ref{energy}) is a three-dimensional spatial integral,
\begin{equation}
\overline{H}=\int \rmd^3\bbox{r}{\mathcal H}(\bbox{r}) ,
\end{equation}
of local energy density ${\mathcal H}(\bbox{r})$ that is a real, scalar,
time-even, and isoscalar function of local densities and their first
and second derivatives. (Isospin-breaking terms, like those
resulting from different neutron and proton masses and from the
Coulomb interaction, can be easily added and, for simplicity, are not
considered in the present study.) In the case of no proton-neutron
mixing, the construction of the most general energy density
that is quadratic in one-body local densities  was
presented in detail in Ref.\ \cite{[Dob96b]}. With the proton-neutron
mixing included, the construction can be performed analogically by
including the additional non-zero local densities derived in Sec.\
\ref{sec2}. Then the energy density can be written in the following
form:
\begin{equation}
{\mathcal H}(\bbox{r}) = \frac{\hbar^{2}}{2m}\tau_0(\bbox{r})
+\sum_{t=0,1}\left(\chi_t(\bbox{r}) +\breve{\chi}_t(\bbox{r})\right),
\label{enden}
\end{equation}
where we assumed that the neutron and proton masses are equal.

The p-h and p-p interaction energy densities, $\chi_t(\bbox{r})$ and
$\breve{\chi}_t$, for $t$=0 depend quadratically on the isoscalar
densities, and for $t$=1 -- on the isovector ones. Based on general rules
of constructing the energy density, Sec.\ \ref{sec2.3}, one obtains
\begin{widetext}
\bnll{chiph}
\label{chi0ph}
{\chi}_{0}(\bbox{r})
&=& C_{0}^{\rho}                            \rho_{0}^{2}\ofbboxofr
+  C_{0}^{\Delta\rho}                      \rho_{0}    \ofbboxofr
\Delta\rho_{0}    \ofbboxofr
+  C_{0}^{\tau}                            \rho_{0}    \ofbboxofr
\tau_{0}    \ofbboxofr
+  C_{0}^{J0}                               {J}_{0}^{2}\ofbboxofr
+  C_{0}^{J1}                          \bbox{J}_{0}^{2}\ofbboxofr
+  C_{0}^{J2}             \underline{\mathsf J}_{0}^{2}\ofbboxofr
+  C_{0}^{\nabla {J}}                      \rho_{0}    \ofbboxofr
\bbox{\nabla}\cdot\bbox{J}_{0}    \ofbboxofr
\nonumber \\
&+& C_{0}^{s}                           \bbox{s}_{0}^{2}\ofbboxofr
+  C_{0}^{\Delta{s}}                   \bbox{s}_{0}    \ofbboxofr
\cdot\Delta\bbox{s}_{0}    \ofbboxofr
+  C_{0}^{T}                           \bbox{s}_{0}    \ofbboxofr
\cdot\bbox{T}_{0}    \ofbboxofr
+  C_{0}^{j}                           \bbox{j}_{0}^{2}\ofbboxofr
+  C_{0}^{\nabla{j}}                   \bbox{s}_{0}    \ofbboxofr
\cdot(\bbox{\nabla}\times\bbox{j}_{0}    \ofbboxofr)
+  C_{0}^{\nabla{s}}(\bbox{\nabla}\cdot\bbox{s}_{0}    \ofbboxofr)^2
+  C_{0}^{F}                           \bbox{s}_{0}    \ofbboxofr
\cdot\bbox{F}_{0}    \ofbboxofr ,
\\ \label{chi1ph}
{\chi}_{1}(\bbox{r})
&=& C_{1}^{\rho}                           \vec{         \rho}^{\,2}\ofbboxofr
+  C_{1}^{\Delta\rho}                     \vec{         \rho}      \ofbboxofr
\circ       \Delta\vec{         \rho}      \ofbboxofr
+  C_{1}^{\tau}                           \vec{         \rho}      \ofbboxofr
\circ             \vec{         \tau}      \ofbboxofr
+  C_{1}^{J0}                             \vec{            J}^{\,2}\ofbboxofr
+  C_{1}^{J1}                             \vec{     \bbox{J}}^{\,2}\ofbboxofr
+  C_{1}^{J2}\underline{                  \vec{ {\mathsf J}}}^{\,2}\ofbboxofr
+  C_{1}^{\nabla {J}}                     \vec{         \rho}      \ofbboxofr
\circ\bbox{\nabla}\cdot\vec{     \bbox{J}}      \ofbboxofr
\nonumber \\
&+& C_{1}^{s}                              \vec{     \bbox{s}}^{\,2}\ofbboxofr
+  C_{1}^{\Delta{s}}                      \vec{     \bbox{s}}      \ofbboxofr
\cdot \circ       \Delta\vec{     \bbox{s}}      \ofbboxofr
+  C_{1}^{T}                              \vec{     \bbox{s}}      \ofbboxofr
\cdot \circ~            \vec{     \bbox{T}}      \ofbboxofr
+  C_{1}^{j}                              \vec{     \bbox{j}}^{\,2}\ofbboxofr
+  C_{1}^{\nabla{j}}                      \vec{     \bbox{s}}      \ofbboxofr
\cdot \circ~(\bbox{\nabla}\times\vec{     \bbox{j}}      \ofbboxofr)
+  C_{1}^{\nabla{s}}
(\bbox{\nabla}\cdot                   \vec{     \bbox{s}}      \ofbboxofr)^2
+  C_{1}^{F}                              \vec{     \bbox{s}}      \ofbboxofr
\cdot \circ~            \vec{     \bbox{F}}      \ofbboxofr,
\enll
where $\times$ stands for the vector product, and
\bnll{chitild}
\label{chitild1}
\breve{\chi}_{0}(\bbox{r})
&=& \breve{C}_{0}^{s}               |\breve{\bbox{s}}_0      \ofbboxofr|^2
+  \breve{C}_{0}^{\Delta{s}}\Re\big(\breve{\bbox{s}}_0^{\,*}\ofbboxofr
\cdot\Delta\breve{\bbox{s}}_0      \ofbboxofr\big)
+  \breve{C}_{0}^{T}        \Re\big(\breve{\bbox{s}}_0^{\,*}\ofbboxofr
\cdot\breve{\bbox{T}}_0      \ofbboxofr\big)
\nonumber \\
&+& \breve{C}_{0}^{j}               |\breve{\bbox{j}}_0      \ofbboxofr|^2
+  \breve{C}_{0}^{\nabla{j}}\Re\big(\breve{\bbox{s}}_0^{\,*}\ofbboxofr
\cdot(\bbox{\nabla}\times\breve{\bbox{j}}_0)     \ofbboxofr\big)
+  \breve{C}_{0}^{\nabla{s}}
|\bbox{\nabla}\cdot\breve{\bbox{s}}_0      \ofbboxofr|^2
+  \breve{C}_{0}^{F}        \Re\big(\breve{\bbox{s}}_0^{\,*}\ofbboxofr
\cdot\breve{\bbox{F}}_0      \ofbboxofr\big),
\\ \label{chitild2}
\breve{\chi}_1(\bbox{r})
&=&  \breve{C}_{1}^{\rho}             |\vec{\breve      {\rho}}      \ofbboxofr|^2
+   \breve{C}_{1}^{\Delta\rho}\Re\big(\vec{\breve      {\rho}}^{\,*}\ofbboxofr
\circ\Delta\vec{\breve      {\rho}}      \ofbboxofr\big)
+   \breve{C}_{1}^{\tau}      \Re\big(\vec{\breve      {\rho}}^{\,*}\ofbboxofr
\circ\vec{\breve      {\tau}}      \ofbboxofr\big)
\nonumber \\
&+&  \breve{C}_{1}^{J0}               |\vec{\breve         {J}}      \ofbboxofr|^2
+   \breve{C}_{1}^{J1}               |\vec{\breve  {\bbox{J}}}      \ofbboxofr|^2
+   \breve{C}_{1}^{J2}|\underline{    \vec{\breve{\mathsf J}}}      \ofbboxofr|^2
+   \breve{C}_{1}^{\nabla{J}} \Re\big(\vec{\breve      {\rho}}^{\,*}\ofbboxofr
\circ\bbox{\nabla}\cdot\vec{\breve  {\bbox{J}}}      \ofbboxofr\big).
\enll
\end{widetext}
In Eqs.\ (\ref{chiph}) and (\ref{chitild}) squares always denote
total lengths in space and/or iso-space, for complex densities taken
in the complex sense, e.g., $|\vec{\breve{\bbox{J}}}(\bbox{r})|^2 =
\sum_{ak}\breve{J}^*_{ak}\breve{J}_{ak}$. In the {\pairing} energy
density (\ref{chitild}) we show only terms in which the products of
real parts are added to products of imaginary parts. According to the
rules based on the $TC$-symmetry, Sec.\ \ref{sec2.3}, similar terms
with both products subtracted from one another are also allowed. We
do not show them explicitly, because they have exactly the form of
Eq.\ (\ref{chitild}), but without complex conjugations and with
absolute values replaced by real parts of products.

When the effective interaction is density-dependent all coupling
constants, $C$'s and $\breve{C}$'s, may also depend on density. If
this is the case, however, terms that can be transformed into one
another by integration by parts are not anymore equivalent. Then,
five more types of terms may appear in the energy density, see Ref.\
\cite{[Dob96b]}; we do not consider such a possibility in the present
study.
Note that in the p-h channel all coupling constants appear in two
flavors, for $t$=0 and 1, while in the p-p channel each one appears
exclusively either for $t$=0, or for $t$=1.

The expression (\ref{enden}) is fairly general. In particular, {\it it
is not based on  any particular two-body  interaction}. However,
if one assumes that
the underlying two-body potential is local and momentum-independent,
the form of  (\ref{enden}) can be simplified and the
number of coupling constants can be reduced. Two particular cases
of practical interest
are discussed in the following.

\subsection{Local gauge invariance}\label{sec3.1}

Under a local gauge transformation \cite{[Dob95e]}, many body
wave function is multiplied by position-dependent phase factor
\begin{equation}\label{eq111}
|\Psi'\rangle = \exp\Big\{i\sum_{j=1}^A\phi(\bbox{r}_j)\Big\}
|\Psi\rangle,
\end{equation}
which induces the following gauge transformations of density matrices
(\ref{rho}) and (\ref{rhobreve}),
\bnll{gauge}
\hat{\rho}'(\bbox{r}st,\bbox{r}'s't')
&=& e^{i\phi(\bbox{r})-i\phi(\bbox{r}')}
\hat{\rho}(\bbox{r}st,\bbox{r}'s't'),
\label{rho-gauge}\\
\hat{\breve{\rho}}'(\bbox{r}st,\bbox{r}'s't')
&=& e^{i\phi(\bbox{r})+i\phi(\bbox{r}')}
\hat{\breve{\rho}}(\bbox{r}st,\bbox{r}'s't').
\label{rhobreve-gauge}
\enll
The Galilean transformation is a local gauge transformation for
$\phi(\bbox{r})$=$\bbox{p}\cdot\bbox{r}$, where $\bbox{p}$
is a constant boost momentum. In analogy to that, one can introduce
the local momentum field defined by
\begin{equation}\label{gauge-mome}
\bbox{p}(\bbox{r}) = \bbox{\nabla}\phi(\bbox{r}) .
\end{equation}

Local and momentum-independent interaction is invariant with respect
to local gauge transformation, and hence energy densities
(\ref{chiph}) and (\ref{chitild}) must then also be independent of the
local gauge. The question whether it is possible to model nuclear
effective interactions in the {\particle} and {\pairing} channels by
a local and momentum-independent interaction, is open. Therefore,
gauge transformation of the energy density can, in principle, be
respected or not, depending on a choice of dynamics one makes.

It is easy to tell when the local energy densities (\ref{chiph})
and (\ref{chitild}) are local-gauge invariant, because properties of
local densities (\ref{scalar-p})--(\ref{tensor-J}) under gauge
transformation read explicitly
\bnll{gauge-dens-ph}
\rho'      _k\ofbboxofr &=& \rho      _k\ofbboxofr
\label{eq112a}  \\
\tau'      _k\ofbboxofr &=& \tau      _k\ofbboxofr
+  2\bbox{p}   \ofbboxofr\cdot\bbox{j}  _k\ofbboxofr
+  \bbox{p}^2  \ofbboxofr\rho           _k\ofbboxofr ,
\label{eq112b}  \\
\bbox{s}'  _k\ofbboxofr &=& \bbox{s}  _k\ofbboxofr ,
\label{eq112c}  \\
\bbox{T}'  _k\ofbboxofr &=& \bbox{T}  _k\ofbboxofr
+  2\bbox{p}   \ofbboxofr\cdot\mathsf{J}_k\ofbboxofr
+  \bbox{p}^2  \ofbboxofr\bbox{s}       _k\ofbboxofr ,
\label{eq112e}  \\
\bbox{j}'  _k\ofbboxofr &=& \bbox{j  }_k\ofbboxofr
+  \bbox{p}    \ofbboxofr\rho           _k\ofbboxofr ,
\label{eq112d}  \\
\bbox{F}'  _k\ofbboxofr &=& \bbox{F}  _k\ofbboxofr
+  \bbox{p}    \ofbboxofr{J}            _k\ofbboxofr
+  \mathsf{J}_k\ofbboxofr\cdot\bbox{p}    \ofbboxofr
+ \bbox{p}    \ofbboxofr\big(\bbox{p}    \ofbboxofr
\cdot\bbox{s}  _k\ofbboxofr\big) ,
\label{eq112g}  \\
\mathsf{J}'_k\ofbboxofr &=& \mathsf{J}_k\ofbboxofr
+  \bbox{p}    \ofbboxofr\otimes\bbox{s}_k\ofbboxofr ,
\label{eq112f}
\enll
where $k$=0,1,2,3, and
\bnll{gauge-dens-pp}
\vec{\breve{\rho~}}'      \ofbboxofr &=& e^{2i\phi               \ofbboxofr}
\vec{\breve{\rho}}      \ofbboxofr
\label{eq113a}  \\
\vec{\breve{\tau~}}'      \ofbboxofr &=& e^{2i\phi               \ofbboxofr}\big(
\vec{\breve{\tau}}      \ofbboxofr
+  i\bbox{p}               \ofbboxofr\cdot\bbox{\nabla}
\vec{\breve{\rho}}      \ofbboxofr
-  \bbox{p}^2              \ofbboxofr
\vec{\breve{\rho}}      \ofbboxofr\big) ,
\label{eq113b}  \\
\breve{\bbox{s}}'  _0\ofbboxofr &=& e^{2i\phi               \ofbboxofr}
\breve{\bbox{s}}      _0\ofbboxofr ,
\label{eq113c}  \\
\breve{\bbox{T}}'  _0\ofbboxofr &=& e^{2i\phi               \ofbboxofr}\big(
\breve{\bbox{T}}      _0\ofbboxofr
+  i(\bbox{p}              \ofbboxofr\cdot\bbox{\nabla})
\breve{\bbox{s}}      _0\ofbboxofr
-  \bbox{p}^2              \ofbboxofr
\breve{\bbox{s}}      _0\ofbboxofr\big) ,
\label{eq113e}  \\
\breve{\bbox{j}}'  _0\ofbboxofr &=& e^{2i\phi               \ofbboxofr}
\breve{\bbox{j}}      _0\ofbboxofr ,
\label{eq113d}  \\
\breve{\bbox{F}}'  _0\ofbboxofr &=& e^{2i\phi               \ofbboxofr}\Big(
\breve{\bbox{F}}      _0\ofbboxofr
+  \tfrac{i}{2}(\bbox{\nabla}\cdot
\breve{\bbox{s}}      _0\ofbboxofr)
\bbox{p}                \ofbboxofr
\nonumber \\ &&~~~~~
+  \tfrac{i}{2}(\bbox{\nabla}\otimes
\breve{\bbox{s}}      _0\ofbboxofr)\cdot
\bbox{p}                \ofbboxofr
-  (\bbox{p}               \ofbboxofr\cdot
\breve{\bbox{s}}      _0\ofbboxofr)
\bbox{p}                \ofbboxofr\Big) ,
\label{eq113g}  \\
\vec{\breve{\mathsf{J}~}}'\ofbboxofr &=& e^{2i\phi               \ofbboxofr}
\vec{\breve{\mathsf{J}}}\ofbboxofr .
\label{eq113f}
\enll
Since all local {\pairing} densities (\ref{gauge-dens-pp}) are
multiplied under the gauge transformation by phase factors
$e^{2i\phi(\bbox{r})}$, products of local {\pairing} densities are
not gauge invariant. Therefore, all terms not shown explicitly in
the {\pairing} energy density [see discussion below Eq.\
(\ref{chitild})] violate the gauge invariance. On the other hand,
products of complex conjugate {\pairing} densities and {\pairing}
densities may be gauge invariant. This obviously is the case for
the pairing, spin, current, and spin-current {\pairing} densities,
while only specific combinations of kinetic, spin-kinetic, and
tensor-kinetic densities are gauge invariant.

Complete list of all {\particle} and {\pairing} gauge-invariant
combinations of local densities reads
\bnll{gauge-chiph}
G_{k}^{\tau}                                    (\bbox{r})
&=& \rho                                 _{k}\ofbboxofr
\tau                                 _{k}\ofbboxofr
-  \bbox{j}                         ^{2}_{k}\ofbboxofr , \\
G_{k}^{T}                                       (\bbox{r})
&=& \bbox{s}                             _{k}\ofbboxofr\cdot
\bbox{T}                             _{k}\ofbboxofr
- {\mathsf J}                       ^{2}_{k}\ofbboxofr \nonumber \\
&=& \bbox{s}                             _{k}\ofbboxofr\cdot
\bbox{T}                             _{k}\ofbboxofr
-  \tthird J                        ^{2}_{k}\ofbboxofr
-  \thalf\bbox{J}                   ^{2}_{k}\ofbboxofr
- \underline{\mathsf J}             ^{2}_{k}\ofbboxofr , \\
G_{k}^{\nabla{J}}                               (\bbox{r})
&=& \rho                                 _{k}\ofbboxofr\bbox{\nabla}\cdot
\bbox{J}                             _{k}\ofbboxofr
+  \bbox{s}                             _{k}\ofbboxofr\cdot(\bbox{\nabla}\times
\bbox{j}                             _{k}\ofbboxofr), \\
G_{k}^{F}                                       (\bbox{r})
&=& \bbox{s}_{k}                             \ofbboxofr\cdot
\bbox{F}_{k}                             \ofbboxofr
-  \thalf \Big(\sum_{a} {\mathsf J}_{kaa}\Big)^{2}   \ofbboxofr
-  \thalf \sum_{ab}{\mathsf J}_{kab} {\mathsf J}_{kba}   \ofbboxofr \nonumber \\
&=& \bbox{s}_{k}                             \ofbboxofr\cdot
\bbox{F}_{k}                             \ofbboxofr
-  \tfrac{2}{3}{J}                  ^{2}_{k}\ofbboxofr
+  \tfrac{1}{4}\bbox{J}             ^{2}_{k}\ofbboxofr
-  \tfrac{1}{2}\underline{\mathsf J}^{2}_{k}\ofbboxofr ,
\enll
where $k$=0,1,2,3, and
\bnll{gauge-chitild}
\breve{G}_{0}^{T}(\bbox{r})
&=&                        \Re\big(\breve{\bbox{s}}_0^{\,*}\ofbboxofr
\cdot\breve{\bbox{T}}_0      \ofbboxofr\big)
-             \tfrac{1}{4}\Re\big(\breve{\bbox{s}}_0^{\,*}\ofbboxofr
\cdot\Delta\breve{\bbox{s}}_0      \ofbboxofr\big), \\
\breve{G}_{0}^{F}(\bbox{r})
&=&                        \Re\big(\breve{\bbox{s}}_0^{\,*}\ofbboxofr
\cdot\breve{\bbox{F}}_0      \ofbboxofr\big)
+  \tfrac{1}{4}|\bbox{\nabla}\cdot\breve{\bbox{s}}_0      \ofbboxofr|^2  , \\
\breve{G}_{k}^{\tau}(\bbox{r})
&=&                        \Re\big(\breve{\rho}    _k^{\,*}\ofbboxofr
\    \breve{\tau}    _k      \ofbboxofr\big)
-             \tfrac{1}{4}\Re\big(\breve{\rho}    _k^{\,*}\ofbboxofr
\Delta\breve{\rho}    _k      \ofbboxofr\big),
\enll
where $k$=1,2,3. Note that terms of the {\pairing} energy density
that depend on $\bbox{\nabla}\times\breve{\bbox{j}}_0$ and
$\bbox{\nabla}\cdot\vec{\breve  {\bbox{J}}}$ are not gauge invariant.

Finally, energy density given by Eqs.\ (\ref{chiph}) and (\ref{chitild})
is gauge invariant provided the coupling constants fulfill the
following constraints,
\bnll{gauge-coup-ph}
C_{t}^{j}        &=& -              C_{t}^{\tau}                       , \\
C_{t}^{J0      } &=& - \tfrac{1}{3} C_{t}^{T} - \tfrac{2}{3} C_{t}^{F} , \\
C_{t}^{J1      } &=& - \tfrac{1}{2} C_{t}^{T} + \tfrac{1}{4} C_{t}^{F} , \\
C_{t}^{J2      } &=& -              C_{t}^{T} - \tfrac{1}{2} C_{t}^{F} , \\
C_{t}^{\nabla j} &=& +              C_{t}^{\nabla J}  ,
\enll
for $t$=0,1, and
\bnll{gauge-coup-pp}
\breve{C}_{0}^{\Delta s}   &=& - \tfrac{1}{4} \breve{C}_{0}^{T}   , \\
\breve{C}_{0}^{\nabla s}   &=& + \tfrac{1}{4} \breve{C}_{0}^{F}   , \\
\breve{C}_{0}^{\nabla j}   &=&   \phantom{-} 0                    , \\
\breve{C}_{1}^{\Delta\rho} &=& - \tfrac{1}{4} \breve{C}_{1}^{\tau}, \\
\breve{C}_{1}^{\nabla J}   &=&   \phantom{-} 0                    .
\enll

\subsection{Skyrme interaction energy functional}\label{sec3.2}

The Skyrme interaction \cite{[Sky59],[Vau72]} is a zero-range local
force that depends on relative momenta up to the second-order. The
complete list of terms giving its matrix element in the
position-spin-isospin representation, including the tensor components
\cite{[Flo75],[Sta77]}, reads
\begin{widetext}
\bn\label{Skyrme}
\hat{V}(\bbox{r}'_1 s'_1t'_1 \bbox{r}'_2s'_2t'_2,\bbox{r}_1s_1t_1\:  \bbox{r}_2s_2t_2)
&=&\Big\{t_0 ( \hat{\delta}^{\sigma} + x_0 \hat{P}^\sigma )
+
\tfrac{1}{6}t_3 ( \hat{\delta}^{\sigma} + x_3 \hat{P}^{\sigma} )
\rho_0^\alpha\left(\thalf(\bbox{r}_1+\bbox{r}_2)\right)
\nonumber \\
&+& \tfrac{1}{2} t_1 ( \hat{\delta}^{\sigma} + x_1 \hat{P}^{\sigma} )
\Big[   \hat{\bbox{k}}^{\prime2}
+\hat{\bbox{k}}^2 \Big]
+ \tfrac{1}{2} t_{\text{e}}
\Big[   {\hat{\bbox{k}}}^{\prime *}\cdot\hat{{\mathsf S}}\cdot{\hat{\bbox{k}}}^{\prime *}
+  \hat{\bbox{k}} \cdot\hat{{\mathsf S}}\cdot\hat{\bbox{k}}  \Big] \\
&+&t_2 ( \hat{\delta}^{\sigma} + x_2 \hat{P}^{\sigma})
{\hat{\bbox{k}}}^{\prime *} \cdot \hat{\bbox{k}}
+ t_{\text{o}}
{\hat{\bbox{k}}}^{\prime *}\cdot\hat{{\mathsf S}} \cdot\hat{\bbox{k}}
+ iW_0 \hat{\bbox{S}}\cdot \Big[
{\hat{\bbox{k}}}^{\prime *} \times \hat{\bbox{k}}\Big]
\Big\}\!\left(\hat{\delta}^{\sigma}\hat{\delta}^{\tau}
- \hat{P}     ^{\sigma}\hat{P}     ^{\tau}P^M\right)
\hat{\delta}_{12},   \nonumber
\en
\end{widetext}
where
\bnll{unit-oper}
\label{unit-sig}
\hat{\delta}^{\sigma}_{s'_1s'_2s_1s_2} &=& \delta_{s'_1s_1}\delta_{s'_2s_2},\\
\label{unit-tau}
\hat{\delta}^{\tau}  _{t'_1t'_2t_1t_2} &=& \delta_{t'_1t_1}\delta_{t'_2t_2},
\enll
and
\bnll{exch-oper}
\label{exch-oper-sig}
\hspace*{-2.5em}
\hat{P}^\sigma_{s_1's_2's_1s_2}\!&=&\!\thalf(\hat{\delta}^{\sigma}_{s'_1s'_2s_1s_2}
\!+\hat{\bbox{\sigma}}_{s_1's_1}\cdot\hat{\bbox{\sigma}}_{s_2's_2})
\!= \delta_{s'_1s_2}\delta_{s'_2s_1} ,\! \\
\label{exch-oper-tau}
\hspace*{-2.5em}
\hat{P}^\tau  _{t_1't_2't_1t_2}\!&=&\!\thalf(\hat{\delta}^{\tau}_{t'_1t'_2t_1t_2}
+\hat{\vec   {\tau}}_{t_1't_1}\circ\hat{\vec   {\tau}}_{t_2't_2})
\!= \delta_{t'_1t_2}\delta_{t'_2t_1} ,
\enll
are the spin and isospin unity and exchange operators, respectively, and
\bnll{spin-oper}
\label{spin-oper-vec}
\hat{\bbox{S}}_{s_1's_2's_1s_2}&=&\hat{\bbox{\sigma}}_{s_1's_1}\delta_{s_2's_2}
+ \hat{\bbox{\sigma}}_{s_2's_2}\delta_{s_1's_1} , \\
\label{spin-oper-ten}
\hat{{\mathsf S}}_{s_1's_2's_1s_2}^{ab}
&=&\tfrac{3}{2}\big(\hat{\bbox{\sigma}}_{s_1's_1}^a
\hat{\bbox{\sigma}}_{s_2's_2}^b
+\hat{\bbox{\sigma}}_{s_1's_1}^b
\hat{\bbox{\sigma}}_{s_2's_2}^a\big)
\nonumber \\
&-&       \delta_{ab}\hat{\bbox{\sigma}}_{s_1's_1}\cdot
\hat{\bbox{\sigma}}_{s_2's_2}
\enll
are two-body vector and tensor spin operators, respectively.
The relative momentum operators,
\bnll{mome-oper}
\label{mome-oper-n}
\hat{\bbox{k}} &=& \tfrac{1}{2i}\left(\bbox{\nabla} _1-\bbox{\nabla} _2\right) , \\
\label{mome-oper-p}
\hat{\bbox{k}}'&=& \tfrac{1}{2i}\left(\bbox{\nabla}'_1-\bbox{\nabla}'_2\right) ,
\enll
act on the delta functions in $\hat{\delta}_{12}$,
\bn
\label{unit-pos}
\hspace*{-1.5em}
\hat{\delta}_{12}(\bbox{r}'_1\bbox{r}'_2,\bbox{r}_1\bbox{r}_2)
&=& \delta(\bbox{r}'_1\!-\!\bbox{r}_1)
\delta(\bbox{r}'_2\!-\!\bbox{r}_2)
\delta(\bbox{r} _1\!-\!\bbox{r}_2)
\nonumber \\
&=& \delta(\bbox{r}'_1\!-\!\bbox{r}_2)
\delta(\bbox{r}'_2\!-\!\bbox{r}_1)
\delta(\bbox{r} _2\!-\!\bbox{r}_1) .
\en
This action has to be understood in the standard sense of derivatives
of distributions.

Whenever the Skyrme interaction (\ref{Skyrme}) is
inserted into integrals, like in Eqs.\
(\ref{hamiltonian})--(\ref{pairpot}), the integration by parts
transfers the derivatives onto appropriate variables in the
remaining parts of integrands.

Numbers $P^M$ are equal to $+1$ or
$-1$ depending on whether in a given term the power of momentum
$\hat{\bbox{k}}$ is even or odd, respectively. Skyrme interaction
written in the form of the integral kernel (\ref{Skyrme}) is
explicitly antisymmetric with respect to exchanging left or right
pairs of variables pertaining to particles 1 and 2.

The Skyrme HFB energy density can be calculated by
inserting the Skyrme interaction (\ref{Skyrme}) directly into
expressions (\ref{partpot}), (\ref{pairpot}), and (\ref{energy}).
Results for the {\particle} channel were published by many authors,
see, e.g., Refs.\ \cite{[Vau72],[Bei75],[Eng75],[Dob95e]}, although
often some terms of interaction (\ref{Skyrme}) were neglected and/or
restricted symmetries were used. Results for the {\pairing} channel
were previously published with tensor terms and the proton-neutron
mixing neglected \cite{[Dob84]}. Here we aim at presenting the
complete set of results.

Calculations leading to expressions for the Skyrme energy density
are tedious, but can be efficiently performed by noting two
simplifying facts. First, the two-body spin operators obey conditions,
\bn
\hat{\bbox{S}}   \hat{P}^\sigma &=& \hat{\bbox{S}}  , \\
\hat{{\mathsf S}}\hat{P}^\sigma &=& \hat{{\mathsf S}} ,
\en
and hence only terms up to linear in spin and isospin Pauli matrices
appear in the antisymmetrized interaction.
Second, the Pauli matrices in (\ref{Skyrme})
pertain to the {\particle} coupling channel, while the momenta -- to
the {\pairing} coupling channel. Hence, calculations may become very
easy once a common, {\particle} or {\pairing}, coupling channel is
used for all the elements of interaction. This requires either
recoupling momenta to the {\particle} channel, or recoupling the
Pauli matrices to the {\pairing} channel. To this end, we separately
consider the {\particle} end {\pairing} energy densities.

\subsubsection{The {\particle} channel}\label{sec3.2.1}

In the {\particle} energy density, indices of the Pauli matrices are
contracted directly with density matrices of particles 1 and 2, and
immediately give non-local densities through appropriate traces in
Eqs.\ (\ref{g1-2})--(\ref{sp1-2}). However, the relative momentum
operators (\ref{mome-oper}) affect both particles at the same time,
and hence have to be first recoupled to forms where the two particles
are acted upon independently, i.e.,
\begin{widetext}
\bnll{mome-recoup}
\thalf\big(\hat{\bbox{k}}^{\prime2}+\hat{\bbox{k}}^2\big)
&=& \tfrac{1}{8}\big(\hat{\bbox{K}}_1^2+    \hat{\bbox{K}}_2^2
-\hat{\bbox{K}}_1  \cdot\hat{\bbox{K}}_2
-4\hat{\bbox{k}}_1  \cdot\hat{\bbox{k}}_2\big)
+  \tfrac{1}{4}\big(\bbox{\nabla}_1\cdot\bbox{\nabla}'_1
+\bbox{\nabla}_2\cdot\bbox{\nabla}'_2\big) , \\
{\hat{\bbox{k}}}^{\prime *} \cdot\hat{\bbox{k}}
&=& \tfrac{1}{8}\big(\hat{\bbox{K}}_1  \cdot\hat{\bbox{K}}_2
-4\hat{\bbox{k}}_1  \cdot\hat{\bbox{k}}_2\big)
+  \tfrac{1}{4}\big(\bbox{\nabla}_1\cdot\bbox{\nabla}'_1
+\bbox{\nabla}_2\cdot\bbox{\nabla}'_2\big) , \\
{\hat{\bbox{k}}}^{\prime *} \times\hat{\bbox{k}}
&=& \tquart\big(\hat{\bbox{K}}_1-\hat{\bbox{K}}_2\big)\times
\big(\hat{\bbox{k}}_2-\hat{\bbox{k}}_1\big) , \\
{\hat{\bbox{k}}}^{\prime *}\otimes{\hat{\bbox{k}}}^{\prime *}+\hat{\bbox{k}}\otimes\hat{\bbox{k}}
&=& \tfrac{1}{4}\big(\hat{\bbox{K}}_1\otimes\hat{\bbox{K}}_1
+\hat{\bbox{K}}_2\otimes\hat{\bbox{K}}_2\big)
-  \tfrac{1}{8}\big(\hat{\bbox{K}}_1\otimes\hat{\bbox{K}}_2
+\hat{\bbox{K}}_2\otimes\hat{\bbox{K}}_1\big)
-  \tfrac{1}{2}\big(\hat{\bbox{k}}_1\otimes\hat{\bbox{k}}_2
+\hat{\bbox{k}}_2\otimes\hat{\bbox{k}}_1\big)
\nonumber \\
&+& \tfrac{1}{4}\big(\bbox{\nabla} _1\otimes\bbox{\nabla}'_1
+\bbox{\nabla} _2\otimes\bbox{\nabla}'_2\big)
+  \tfrac{1}{4}\big(\bbox{\nabla}'_1\otimes\bbox{\nabla} _1
+\bbox{\nabla}'_2\otimes\bbox{\nabla} _2\big) , \\
{\hat{\bbox{k}}}^{\prime *} \otimes\hat{\bbox{k}} +\hat{\bbox{k}}\otimes{\hat{\bbox{k}}}^{\prime *}
&=& \tfrac{1}{8}\big(\hat{\bbox{K}}_1\otimes\hat{\bbox{K}}_2
+\hat{\bbox{K}}_2\otimes\hat{\bbox{K}}_1\big)
-  \tfrac{1}{2}\big(\hat{\bbox{k}}_1\otimes\hat{\bbox{k}}_2
+\hat{\bbox{k}}_2\otimes\hat{\bbox{k}}_1\big)
\nonumber \\
&+& \tfrac{1}{4}\big(\bbox{\nabla} _1\otimes\bbox{\nabla}'_1
+\bbox{\nabla} _2\otimes\bbox{\nabla}'_2\big)
+  \tfrac{1}{4}\big(\bbox{\nabla}'_1\otimes\bbox{\nabla} _1
+\bbox{\nabla}'_2\otimes\bbox{\nabla} _2\big) ,
\enll
\end{widetext}
where
\bnll{mome-1-2}
\hat{\bbox{k}}_1 &=&
\tfrac{1}{2i}\left(\bbox{\nabla}_1-\bbox{\nabla}'_1\right) , \\
\hat{\bbox{k}}_2 &=&
\tfrac{1}{2i}\left(\bbox{\nabla}_2-\bbox{\nabla}'_2\right) ,
\enll
and
\bnll{mome-1+2}
\hat{\bbox{K}}_1 &=&
-i\left(\bbox{\nabla}_1+\bbox{\nabla}'_1\right) , \\
\hat{\bbox{K}}_2 &=&
-i\left(\bbox{\nabla}_2+\bbox{\nabla}'_2\right) .
\enll
Final results can now be easily obtained by noting that relative
momenta (\ref{mome-1-2}) lead to the current densities
(\ref{vector-j-ph}) and (\ref{tensor-J-ph}), total momenta
(\ref{mome-1+2}) lead to derivatives of local densities, and the
scalar and tensor products of individual momenta lead to kinetic
densities (\ref{scalar-k-ph}), (\ref{vector-T-ph}), and
(\ref{vector-F-ph}).

The zero-order (density-dependent) {\particle} coupling constants of the
energy density (\ref{chiph}) are expressed by the
Skyrme force parameters as
\bnll{zero-order-ph}
C_{0}^{\rho} &=& \pr\tfrac{3}{8} t_0
+ \tfrac{3}{48}t_3        \rho_0^\alpha(\bbox{r}) , \\
C_{0}^{s}    &=& \pr\tfrac{1}{8} t_0(2x_0-1)
+ \tfrac{1}{48}t_3(2x_3-1)\rho_0^\alpha(\bbox{r}) , \\
C_{1}^{\rho} &=&  - \tfrac{1}{8} t_0(2x_0+1)
- \tfrac{1}{48}t_3(2x_3+1)\rho_0^\alpha(\bbox{r}) ,\\
C_{1}^{s}    &=&  - \tfrac{1}{8} t_0
- \tfrac{1}{48}t_3        \rho_0^\alpha(\bbox{r}) ,
\enll
and the second-order coupling constants are given in Table
\ref{tab3}. One can immediately see that the gauge-invariance
conditions (\ref{gauge-coup-ph}) are fulfilled. This is so because
the momentum-dependent terms of the Skyrme interaction obey the
Galilean invariance \cite{[Eng75],[Dob95e]}

Since seven Skyrme force parameters define twenty four second-order
{\particle} coupling constants, in the resulting Skyrme energy
density there is a high degree of dependency. First, as is well-known
\cite{[Rei95]}, a single spin-orbit parameter $W_0$ determines all
four spin-orbit coupling constants $C_t^{\nabla{J}}$ and
$C_t^{\nabla{j}}$, for $t$=0 and 1. Second, four Skyrme parameters,
$t_1$, $x_1$, $t_2$, and $x_2$, uniquely determine four coupling
constants $C_t^{\Delta\rho}$ and $C_t^{\tau}$, for $t$=0 and 1.
Third, two tensor Skyrme parameters, $t_e$ and $t_o$, uniquely
determine either isoscalar or isovector coupling constants,
$C_t^{\nabla{s}}$ and $C_t^{F}$. Once such a r\^ole of the seven
Skyrme parameters is fixed, values of the remaining coupling
constants are also uniquely fixed.

\begin{table}
\caption[T3]{Second-order coupling constants of the {\particle} energy density
(\protect\ref{chiph}) as functions of parameters of the Skyrme
interaction (\protect\ref{Skyrme}), expressed by the formula:
$C=\frac{A}{192}(at_1+bt_1x_1+ ct_2+dt_2x_2+et_{\text{e}}+ft_{\text{o}}+gW_0)$.}
\label{tab3}
\begin{ruledtabular}
\begin{tabular}{l|rcrrrrrrr}
&   $A$ &&  $a$ &  $b$ &  $c$ &  $d$ &  $e$ &  $f$ &  $g$ \\
\hline
$C_0^{\Delta\rho} $ &     3 && $-$9 &    0 & \ph5 & \ph4 &    0 &    0 &    0 \\
$C_0^{\tau}       $ &    12 &&    3 &    0 &    5 &    4 &    0 &    0 &    0 \\
$C_0^{J0}         $ &  $-$4 && $-$1 &    2 &    1 &    2 &   10 &   30 &    0 \\
$C_0^{J1}         $ &  $-$6 && $-$1 &    2 &    1 &    2 & $-$5 &$-$15 &    0 \\
$C_0^{J2}         $ & $-$12 && $-$1 &    2 &    1 &    2 &    1 &    3 &    0 \\
$C_0^{\nabla{J}}  $ &    48 &&    0 &    0 &    0 &    0 &    0 &    0 & $-$3 \\
\hline
$C_0^{\Delta s}   $ &     3 &&    3 & $-$6 &    1 &    2 &    6 & $-$6 &    0 \\
$C_0^{T}          $ &    12 && $-$1 &    2 &    1 &    2 & $-$2 & $-$6 &    0 \\
$C_0^{j}          $ & $-$12 &&    3 &    0 &    5 &    4 &    0 &    0 &    0 \\
$C_0^{\nabla{s}}  $ &    18 &&    0 &    0 &    0 &    0 &    3 & $-$3 &    0 \\
$C_0^{F}          $ &    72 &&    0 &    0 &    0 &    0 &    1 &    3 &    0 \\
$C_0^{\nabla{j}}  $ &    48 &&    0 &    0 &    0 &    0 &    0 &    0 & $-$3 \\
\hline
$C_1^{\Delta\rho} $ &     3 &&    3 &    6 &    1 &    2 &    0 &    0 &    0 \\
$C_1^{\tau}       $ &    12 && $-$1 & $-$2 &    1 &    2 &    0 &    0 &    0 \\
$C_1^{J0}         $ &  $-$4 && $-$1 &    0 &    1 &    0 &$-$10 &   10 &    0 \\
$C_1^{J1}         $ &  $-$6 && $-$1 &    0 &    1 &    0 &    5 & $-$5 &    0 \\
$C_1^{J2}         $ & $-$12 && $-$1 &    0 &    1 &    0 & $-$1 &    1 &    0 \\
$C_1^{\nabla{J}}  $ &    48 &&    0 &    0 &    0 &    0 &    0 &    0 & $-$1 \\
\hline
$C_1^{\Delta s}   $ &     3 &&    3 &    0 &    1 &    0 & $-$6 & $-$2 &    0 \\
$C_1^{T}          $ &    12 && $-$1 &    0 &    1 &    0 &    2 & $-$2 &    0 \\
$C_1^{j}          $ & $-$12 && $-$1 & $-$2 &    1 &    2 &    0 &    0 &    0 \\
$C_1^{\nabla{s}}  $ &    18 &&    0 &    0 &    0 &    0 & $-$3 & $-$1 &    0 \\
$C_1^{F}          $ &    72 &&    0 &    0 &    0 &    0 & $-$1 &    1 &    0 \\
$C_1^{\nabla{j}}  $ &    48 &&    0 &    0 &    0 &    0 &    0 &    0 & $-$1 \\
\end{tabular}
\end{ruledtabular}
\end{table}

\subsubsection{The {\pairing} channel}\label{sec3.2.2}

In the {\pairing} energy density, each operator of the relative
momentum, $\hat{\bbox{k}}'$ and $\hat{\bbox{k}}$, acts on variables
of {\it the same} density matrix, and thus no recoupling is necessary.
Terms of the interaction that are linear in momenta then lead to
current densities (\ref{vector-j-pp}) and (\ref{tensor-J-pp}), while terms
that are quadratic in momenta lead to derivatives of local densities
and to kinetic densities (\ref{scalar-k-pp}), (\ref{vector-T-pp}), and
(\ref{vector-F-pp}), because
\bnll{ktimesk}
\hat{\bbox{k}}^2 &=&
-\tquart\big(\bbox{\nabla}_1 + \bbox{\nabla}_2\big)^2
+\bbox{\nabla}_1\cdot\bbox{\nabla}_2 , \\
\hat{\bbox{k}}\otimes\hat{\bbox{k}}& = &
-\tquart\big(\bbox{\nabla}_1 + \bbox{\nabla}_2\big)\otimes\big(\bbox{\nabla}_1 + \bbox{\nabla}_2\big)
\nonumber \\
&&+\thalf\big(\bbox{\nabla}_1\otimes\bbox{\nabla}_2+\bbox{\nabla}_2\otimes\bbox{\nabla}_1\big) .
\enll

However, in the {\pairing} energy density, indices of Pauli matrices
couple together the two density matrices, and hence do require recoupling
to the {\pairing} channel. These recoupling formulae can be obtained by means of
the standard algebra of angular momentum. A sum of the three Clebsch-Gordan
coefficients appropriate to the present case
reads \cite{[Var88]}
\begin{widetext}
\bn\label{CGsum}
&& 4s'_2s_2  \sum_{\mu_1\mu_2}\langle \thalf s_1\lambda_1\mu_1|\thalf s'_1\rangle
\langle\thalf -\! s_2\lambda_2\mu_2|\thalf -\! s'_2\rangle
\langle\lambda_1-\! \mu_1\lambda_2-\! \mu_2|\lambda\mu\rangle
\nonumber \\
&&~~~~~~=\sum_{l'm',lm}(-1)^{\lambda_1-\lambda_2 +l'}(2l'+1)(2l+1)
\left\{\ba{ccc}\thalf & \thalf & \lambda_1 \\
\thalf & \thalf & \lambda_2 \\
l' & l & \lambda \ea \right\}
\times \langle \thalf s'_2l'm'|\thalf s'_1\rangle
\langle \thalf s_1lm|\thalf s_2\rangle
\langle l'-\! m'l-\! m|\lambda\mu \rangle
\en
Taking relevant combinations of $\lambda_1,\, \lambda_2 =0,\, 1$ and
$\lambda =0,\, 1,\, 2$, one obtains:
\bnll{spin-coup}
4s'_2s_2\hat{\delta}^{\sigma}_{s'_1,-s'_2-s_1,-s_2}
&=&    \pr\thalf \delta_{s'_2s'_1}\delta_{s_2s_1}
+  \thalf \bbox{\sigma}^*_{s'_2s'_1}\cdot\bbox{\sigma}_{s_2s_1}     ,
\label{spin-unit}\\
4s'_2s_2\hat{P}^\sigma_{s'_1,-s'_2s_1,-s_2}
&=&    -  \thalf \delta_{s'_2s'_1}\delta_{s_2s_1}
+  \thalf \bbox{\sigma}^*_{s'_2s'_1}\cdot\bbox{\sigma}_{s_2s_1}     ,
\label{spin-exch}\\
4s'_2s_2\hat{\bbox{S}}_{s'_1,-s'_2s_1,-s_2}
&=& -i\,\bbox{\sigma}^*_{s'_2s'_1}\times\bbox{\sigma}_{s_2s_1} ,
\\
4s'_2s_2\hat{{\mathsf S}}^{ab}_{s'_1,-s'_2s_1,-s_2}
&=& -\tfrac{3}{2}\big(\bbox{\sigma}^{a*}_{s'_2s'_1}\bbox{\sigma}^b_{s_2s_1}
+\bbox{\sigma}^{b*}_{s'_2s'_1}\bbox{\sigma}^a_{s_2s_1}\big)
+\,\delta_{ab}\bbox{\sigma}^*_{s'_2s'_1}\cdot\bbox{\sigma}_{s_2s_1}
\enll
\end{widetext}
and the formulae similar to Eqs.\ (\ref{spin-unit}) and (\ref{spin-exch})
are obtained for $\hat{\delta}^{\tau}$ and $\hat{P}^{\tau}$, respectively.

The two zero-order (density-dependent) {\pairing} coupling constants
of the energy density (\ref{chitild}) are related to the Skyrme
parameters in the following way:
\bnll{Skypair}
\breve{C}_{0}^{s}    &=& \tfrac{1}{8} t_0(1+x_0)
+  \tfrac{1}{48}t_3(1+x_3)\rho_0^{\alpha}(\bbox{r}),\\
\breve{C}_{1}^{\rho} &=& \tfrac{1}{8} t_0(1-x_0)
+  \tfrac{1}{48}t_3(1-x_3)\rho_0^{\alpha}(\bbox{r}),
\enll
and the second-order {\pairing} coupling constants are given in Table
\ref{tab4}. Similar to the p-h case,  the gauge-invariance conditions
(\ref{gauge-coup-pp}) are met.

\begin{table}
\caption[T3]
{Second-order coupling constants of the {\pairing} energy density
(\protect\ref{chitild}) as functions of parameters of the Skyrme
interaction (\protect\ref{Skyrme}), expressed by the formula:
$\breve{C}=\frac{A}{96}(at_1+bt_1x_1+ ct_2+dt_2x_2+et_{\text{e}}+ft_{\text{o}}+gW_0)$.}
\label{tab4}
\begin{ruledtabular}
\begin{tabular}{l|rcrrrrrrr}
&  $A$ && $a$ & $b$ & $c$ & $d$ & $e$ & $f$ & $g$ \\
\hline
$\breve{C}_0^{\Delta s}  $& $-$3 &&   1 &   1 &\ph0 &   0 & 2 &   0 &   0 \\
$\breve{C}_0^{T}         $&   12 &&   1 &   1 &   0 &   0 &   2 &   0 &   0 \\
$\breve{C}_0^{j}         $&   12 &&   0 &   0 &   1 &$-$1 &   0 &   0 &   0 \\
$\breve{C}_0^{\nabla{s}} $&$-$18 &&   0 &   0 &   0 &   0 &   1 &   0 &   0 \\
$\breve{C}_0^{F}         $&$-$72 &&   0 &   0 &   0 &   0 &   1 &   0 &   0 \\
$\breve{C}_0^{\nabla{j}} $&    0 &&   0 &   0 &   0 &   0 &   0 &   0 &   0 \\
\hline
$\breve{C}_1^{\Delta\rho}$& $-$3 &&   1 &$-$1 &   0 &   0 &   0 &   0 &   0 \\
$\breve{C}_1^{\tau}      $&   12 &&   1 &$-$1 &   0 &   0 &   0 &   0 &   0 \\
$\breve{C}_1^{J0}        $&    4 &&   0 &   0 &   1 &   1 &   0 &$-$10&   4 \\
$\breve{C}_1^{J1}        $&    6 &&   0 &   0 &   1 &   1 &   0 &   5 &   2 \\
$\breve{C}_1^{J2}        $&   12 &&   0 &   0 &   1 &   1 &   0 &$-$1 &$-$2 \\
$\breve{C}_1^{\nabla{J}} $&    0 &&   0 &   0 &   0 &   0 &   0 &   0 &   0 \\
\end{tabular}
\end{ruledtabular}
\end{table}

Equivalently, the density-dependent zero-range pairing force
$V_{\text{pair}}$ can be used in the p-p channel
\cite{[Boc67],[Cha76],[Kad78],[Dob01a]},
\begin{equation}\label{ddpi}
V_{\text{pair}}(\bbox{r},\bbox{r}') =
f_{\text{pair}}(\bbox{r})\delta(\bbox{r}-\bbox{r}'),
\end{equation}
for
\begin{equation}
f_{\text{pair}}(\bbox{r}) =
V_0\left\{1+x_0\hat{P}^\sigma
-\left[\frac{\rho_0(\bbox{r})}{\rho_c}\right]^\alpha
(1+x_3\hat{P}^\sigma)\right\} ,
\end{equation}
where $\hat{P}^\sigma$ is the spin-exchange operator  (\ref{exch-oper-sig}).
In such a case, coupling constants (\ref{Skypair}) read
\bnll{Delpair}
\breve{C}_{0}^{s}    &=& \tfrac{1}{8} V_0(1+x_0)
-  \tfrac{1}{8} V_0(1+x_3)
\left[\frac{\rho_0(\bbox{r})}{\rho_c}\right]^\alpha ,\\
\breve{C}_{1}^{\rho} &=& \tfrac{1}{8} V_0(1-x_0)
-  \tfrac{1}{8} V_0(1-x_3)
\left[\frac{\rho_0(\bbox{r})}{\rho_c}\right]^\alpha .
\enll
Note that when only the isovector pairing is used, as in most LDA
applications to date, the exchange parameters $x_0$ and $x_3$ are
redundant in the definition of the isovector coupling constant
$\breve{C}_{1}^{\rho}$, and hence are usually set to 0. However, if one
wants to independently model the isoscalar and isovector pairing
intensity, one has to use non-zero values of $x_0$ and $x_3$.

For the Gogny interaction \cite{[RS80]}, the zero-range
den\-si\-ty-dependent term $t_3$ with $\alpha$=1/3 was used in order to
enforce proper saturation properties. The corresponding exchange
parameter $x_3$=1 was used to prevent this zero-range force from
contributing to the isovector pairing channel. However, such a
choice, when  applied literally  to the proton-neutron mixing case,
might lead to a very strong repulsive isoscalar pairing interaction.

The term of $\breve{\chi}$ coming from the spin-orbit interaction
contains the combination of components of the {\pairing} spin-current
density $\vec{\breve{\mathsf J}}$,

\begin{equation}
\sum_{ab}\left(
 \vec{\breve{\mathsf J}}^{\ast}_{aa}\circ\vec{\breve{\mathsf J}}_{bb}
-\vec{\breve{\mathsf J}}^{\ast}_{ab}\circ\vec{\breve{\mathsf J}}_{ba}\right)
= \tfrac{2}{3}|\vec{\breve{      {{J}}}}|^2
+ \thalf      |\vec{\breve{ {\bbox{J}}}}|^2
- |\underline{ \vec{\breve{\mathsf J}}} |^2 ,
\end{equation}
that is different from that coming from the tensor $t_{\text{o}}$ term,
\bn &&
\sum_{ab}\left(\vec{\breve{\mathsf J}}^{\ast}_{ab}\circ\vec{\breve{\mathsf J}}_{ab}
-\tfrac{3}{2}  \vec{\breve{\mathsf J}}^{\ast}_{aa}\circ\vec{\breve{\mathsf J}}_{bb}
-\tfrac{3}{2}  \vec{\breve{\mathsf J}}^{\ast}_{ab}\circ\vec{\breve{\mathsf J}}_{ba}\right)
~~~~~~~~~~~~~~~~~~\nonumber \\ &&~~~~~~~~~~~~~~~~~~~~~~~~
= - \tfrac{5}{3}      |\vec{\breve{      {{J}}}}|^2
+   \tfrac{5}{4}      |\vec{\breve{ {\bbox{J}}}}|^2
-   \thalf|\underline{ \vec{\breve{\mathsf J}}} |^2 ,
\en
and from that coming from the central $t_2$ term,
\begin{equation}
|\vec{\breve{\mathsf J}}|^2 =
\sum_{ab}\left(\vec{\breve{\mathsf J}}^{\ast}_{ab}\circ\vec{\breve{\mathsf J}}_{ab}\right)
= \tthird     |\vec{\breve{      {{J}}}}|^2
+ \thalf      |\vec{\breve{ {\bbox{J}}}}|^2
+ |\underline{ \vec{\breve{\mathsf J}}} |^2 .
\end{equation}
Therefore, by setting appropriate values of the $t_2(1+x_2)$, $W_0$,
and $t_{\text{o}}$ parameters, one can obtain arbitrary values of the
spin-current coupling constants $\breve{C}_1^{J0}$,
$\breve{C}_1^{J1}$, and $\breve{C}_1^{J2}$. Similarly, parameter
$t_2(1-x_2)$ allows for fixing an arbitrary value of the current coupling
constant $\breve{C}_0^{j}$. On the other hand, parameter
$t_1(1+x_1)$ defines two isoscalar coupling constants,
$\breve{C}_0^{\Delta s}$ and $\breve{C}_0^{T}$, parameter
$t_{\text{e}}$ defines another two isoscalar coupling constants,
$\breve{C}_0^{\nabla{s}}$ and $\breve{C}_0^{F}$, and parameter
$t_1(1-x_1)$ defines two isovector coupling constants,
$\breve{C}_1^{\Delta\rho}$ and $\breve{C}_1^{\tau}$; hence, these
pairs of coupling constants are not independent from one another.
These three pairs of dependencies reflect, in fact, the three
gauge invariance conditions (\ref{gauge-chitild}).
In this way, seven Skyrme force parameters determine ten coupling constants
in the {\pairing} channel.
Finally, the Skyrme interaction does not give any non-zero values for
the spin-orbit coupling constants $\breve{C}_0^{\nabla{j}}$ and
$\breve{C}_1^{\nabla{J}}$. Therefore, up to the gauge invariance
conditions, the Skyrme interaction fully determines the energy
density in the {\pairing} channel.

\section{The {\Particle} and {\Pairing} Mean Fields}\label{sec6}

By varying the energy functional (\ref{energy}) with respect to the
density matrices one obtains the {\particle} and {\pairing} mean
field Hamiltonians,
\begin{widetext}
\bnll{varall}
\hat{h}(\bbox{r}'s't',\bbox{r}st)
&=& \frac{\delta \overline{H}[\hat{\rho},\hat{\breve{\rho}},\hat{\breve{\rho}}^+]}
{\delta \hat{\rho}(\bbox{r}st,\bbox{r}'s't')}
= - \frac{\hbar^{2}}{2m}\delta(\bbox{r}-\bbox{r}')
\bbox{\nabla}\cdot  \bbox{\nabla}\,\delta_{s's}\delta_{t't}
+   \hat{\Gamma}              (\bbox{r}'s't',\bbox{r}st)
+   \hat{\Gamma}_{\text{r}}(\bbox{r}'s't',\bbox{r}st) ,
\label{var} \\
\hat{\breve{h}}(\bbox{r}'s't',\bbox{r}st)
&=& \frac{\delta \overline{H}[\hat{\rho},\hat{\breve{\rho}},\hat{\breve{\rho}}^+]}
{\delta \hat{\breve{\rho}}^+(\bbox{r}st,\bbox{r}'s't')}
= \hat{\breve{\Gamma}}              (\bbox{r}'s't',\bbox{r}st)
+ \hat{\breve{\Gamma}}_{\text{r}}(\bbox{r}'s't',\bbox{r}st) .
\label{vartild}
\enll
\end{widetext}
The rearrangement potentials $\hat{\Gamma}_{\text{r}}$ and
$\hat{\breve{\Gamma}}_{\text{r}}$ result from the density
dependence of effective interactions on the {\particle} and
{\pairing} densities, respectively. Usually effective interactions
are assumed to depend only on the {\particle} density matrix (most
often, only on the isoscalar particle density $\rho_0$). In that case
the {\pairing} rearrangement potential vanishes. However, one cannot
forget that the dependence of the p-p interaction on the particle
density results in a corresponding contribution to the p-h
rearrangement potential. In what follows, to simplify the
presentation we do not show the rearrangement terms explicitly.

Within the LDA, the mean-field Hamiltonians being originally, like the Skyrme interaction
of Eq.\ (\ref{Skyrme}), either distributions or derivatives of distributions, can,
when acting as the integral kernels, be expressed as  local, momentum
dependent operators, i.e.,
\bnll{eq81}
\hat      {{h}}(\bbox{r}'s't',\bbox{r}st)
&=&\delta (\bbox{r}-\bbox{r}')\hat      {{h}}(\bbox{r};s't',st), \\
\hat{\breve{h}}(\bbox{r}'s't',\bbox{r}st)
&=&\delta (\bbox{r}-\bbox{r}')\hat{\breve{h}}(\bbox{r};s't',st).
\enll
The kinetic energy term in Eq.\ (\ref{var}) is already expressed in such a form.
The mean-fields Hamiltonians
are the second-order operators in momentum
and matrices in the spin and isospin spaces. The
isospin structure of the local p-h and p-p mean-field Hamiltonians
reads
\bnll{eq82}
\hspace*{-1.5em}
\hat{h}(\bbox{r};s't',st) &=&
{h}_0(\bbox{r};s',s)\delta_{t't}+%
\vec{h}(\bbox{r};s',s)\circ \hat{\vec{\tau}}_{t't} , \\
\hspace*{-1.5em}
\hat{\breve{h}}(\bbox{r};s't',st) &=&
\breve{h}_0(\bbox{r};s',s)\delta_{t't}+%
\vec{\breve{h}}(\bbox{r};s',s)\circ \hat{\vec{\tau}}_{t't},
\enll
respectively.
The isoscalar and isovector parts of the {\particle} mean-field Hamiltonian
can be presented  in the compact form
\begin{widetext}
\bn
h_k(\bbox{r};s',s)
&=&- \frac{\hbar^{2}}{2m}\bbox{\nabla}^2\delta_{s's}\delta_{k0}
+ U_k\ofbboxofr\delta_{s's}
+ \bbox{\Sigma}_k\ofbboxofr\cdot\hat{\bbox{\sigma}}_{s's}
+ \tfrac{1}{2i}\big[\bbox{I}_k\ofbboxofr\delta_{s's}
+ ({\mathsf B}_k\ofbboxofr\cdot\hat{\bbox{\sigma}}_{s's})\big]\cdot\bbox{\nabla}
+ \tfrac{1}{2i}\bbox{\nabla}\cdot\big[\bbox{I}_k\ofbboxofr\delta_{s's}
+ ({\mathsf B}_k\ofbboxofr\cdot\hat{\bbox{\sigma}}_{s's})\big]
\nonumber \\ &&
- \bbox{\nabla}\cdot\big[M_k\ofbboxofr\delta_{s's}
+ \bbox{C}_k\ofbboxofr\cdot\hat{\bbox{\sigma}}_{s's}\big]\bbox{\nabla}
- \bbox{\nabla}\cdot \bbox{D}_k\ofbboxofr\,\hat{\bbox{\sigma}}_{s's}\cdot \bbox{\nabla}
\label{ha}
\en
\end{widetext}
for $k=0,1,2,3$, and where
\begin{equation}\label{tenact}
({\mathsf B}\cdot\hat{\bbox{\sigma}})_a
=\sum_b{\mathsf B}_{ab}\hat{\bbox{\sigma}}^b ,
\end{equation}
for $a=x,y,z$, is the $a$'th component of a space vector. The names
of symbols are inspired by those introduced in Ref.\ \cite{[Eng75]}.
Since the {\particle} density matrix is hermitian, the {\particle}
mean-field Hamiltonian is also hermitian and, thus, all the
potentials, $M_k$, $U_k$, ${\mathsf B}_k$, $\bbox{C}_k$,
$\bbox{D}_k$, $\bbox{I}_k$, and $\bbox{\Sigma}_k$ are real.

The general form of the mean-field Hamiltonian (\ref{ha}) can be
constructed from the momentum $-i\bbox{\nabla}$ and spin
$\hat{\bbox{\sigma}}$ operators, based only on the symmetry
properties. Apart from the one-body kinetic energy [the first term in
Eq.\ (\ref{ha})], the expansion in momentum gives: (i) zero-order
terms with scalar ($U_k$) and pseudovector ($\bbox{\Sigma}_k$)
potentials, (ii) first-order terms with vector ($\bbox{I}_k$) and
pseudotensor (${\mathsf B}_k$) potentials, (iii) second-order-scalar
terms with scalar ($M_k$) and pseudoscalar ($\bbox{C}_k$) effective
masses, and (iv) second-order-tensor terms. In principle, the most
general form of the last category should involve tensor and
third-order-pseudotensor potentials. However, in Eq.\ (\ref{ha}) we
show only the particular form of it that corresponds to the energy
density (\ref{chiph}).

According to Eqs.\ (\ref{varall}) the  p-h mean-field Hamiltonian is the functional derivative
of the energy functional over  the hermitian p-h density matrix. Functional derivatives of integrals
of type:
\begin{equation}\label{denint}
\overline{\langle f\rho   \rangle}= \int \delta (\bbox{r}_1-\bbox{r}_2)f(\bbox{r}_1)\rho(\bbox{r}_1,\bbox{r}_2)
\rmd^3\bbox{r}_1\rmd^3\bbox{r}_2,
\end{equation}
where function $f$ is treated as independent of densities and $\rho$ represents a p-h non-local density,
can easily be calculated using Eqs.\ (\ref{g1-2}) and (\ref{s1-2}). Bearing in mind that
\begin{widetext}
\bn\label{fuderrho}
\frac{\delta \hat{\rho}(\bbox{r}_1s_1t_1,\bbox{r}_2s_2t_2)}{\delta\hat{\rho}(\bbox{r}st,\bbox{r}'s't')}
&=&\delta(\bbox{r}_1-\bbox{r})\delta(\bbox{r}_2-\bbox{r}')\delta_{s_1s}\delta_{s_2s'}\delta_{t_1t}\delta_{t_2t'},
\nonumber \\
\en
one has
\bnll{fuder}
\frac{\delta \overline{\langle f\rho _k\rangle}}{\delta\hat{\rho}(\bbox{r}st,\bbox{r}'s't')}&=&
\delta(\bbox{r}'-\bbox{r})f(\bbox{r})\delta_{s's}\hat{\tau}^k_{t't}, \\
\frac{\delta\overline{\langle f\bbox{s}_k\rangle}}{\delta\hat{\rho}(\bbox{r}st,\bbox{r}'s't')}&=&
\delta(\bbox{r}'-\bbox{r})f(\bbox{r})\hat{\bbox{\sigma}}_{s's}\hat{\tau}^k_{t't}
\enll
for $k=0,1,2,3$. The functional derivatives of integrals of local differential densities are obtained from
Eqs.\ (\ref{fuder}) through integration by parts. Then, the functional derivatives become dependent
on derivatives of the Dirac delta function and thus, in accordance with Eqs.\ (\ref{eq81}), again act as local
differential operators.
They read:
\bnll{fuderdiff}
\frac{\delta \overline{\langle f\bbox{j}_k\rangle}}{\delta\hat{\rho}(\bbox{r}st,\bbox{r}'s't')}
&=&
\tfrac{1}{2i}\delta(\bbox{r}'-\bbox{r})\big(\bbox{\nabla}f(\bbox{r})
+f(\bbox{r})\bbox{\nabla}\big)
\delta_{s's}\hat{\tau}^k_{t't},  \\
\frac{\delta\overline{\langle f\mathsf{J}_{kab}\rangle}}{\delta\hat{\rho}(\bbox{r}st,\bbox{r}'s't')}
&=&
\tfrac{1}{2i}\delta(\bbox{r}'-\bbox{r})\big(\bbox{\nabla}_af(\bbox{r})
+f(\bbox{r})\bbox{\nabla}_a\big)
\hat{\bbox{\sigma}}^b_{s's}\hat{\tau}^k_{t't},
\enll
\bnll{fuderdiff2}
\hspace*{-1.5em}
\frac{\delta\overline{\langle f\bbox{\nabla}_a\bbox{\nabla}^{\prime}_b
\rho _k\rangle}}{\delta\hat{\rho}(\bbox{r}st,\bbox{r}'s't')}&=
-\delta(\bbox{r}'-\bbox{r})&\bbox{\nabla}_af(\bbox{r})\bbox{\nabla}_b
\delta_{s's}\hat{\tau}^k_{t't}, \\
\hspace*{-1.5em}
\frac{\delta\overline{\langle f\bbox{\nabla}_a\bbox{\nabla}^{\prime}_b\bbox{s}_{kc}\rangle}}{\delta\hat{\rho}(\bbox{r}st,\bbox{r}'s't')}&=
-\delta(\bbox{r}'-\bbox{r})&\bbox{\nabla}_af(\bbox{r})\bbox{\nabla}_b
\hat{\bbox{\sigma}}^c_{s's}\hat{\tau}^k_{t't}
\enll
for $k=0,1,2,3$ and $a,b,c=x,y,z$.
Calculations
of the functional derivatives over the density matrix are equivalent
to the rules for variations over single-particle wavefunctions given
by Engel {\it et al.} \cite{[Eng75]}.
Using formulae given above, Eqs.\ (\ref{fuder})--(\ref{fuderdiff2}),
one obtains the following relations between the
potentials defining the p-h mean field (\ref{ha}) and the
local {\particle} densities defining the energy density (\ref{chiph}),
\bnll{potall}
U_k(\bbox{r})&=&2C^{\rho}_t\rho_k\ofbboxofr
+2C^{\Delta \rho}_t\Delta \rho_k\ofbboxofr
+ C^{\tau}_t\tau_k\ofbboxofr
+ C^{\nabla J}_t\bbox{\nabla}\cdot
\bbox{J}_k\ofbboxofr, \\
\bbox{\Sigma}_k(\bbox{r})&=&2C^s_t\bbox{s}_k\ofbboxofr
+ 2(C^{\Delta s}_t-C^{\nabla s}_t)\Delta
\bbox{s}_k\ofbboxofr
- 2C^{\nabla s}_t\bbox{\nabla}\times
(\bbox{\nabla}\times\bbox{s}_k\ofbboxofr)
+C^T_t\bbox{T}_k\ofbboxofr
+ C^F_t\bbox{F}_k\ofbboxofr
+ C^{\nabla j}_t\bbox{\nabla}\times
\bbox{j}_k\ofbboxofr, \\
\bbox{I}_k(\bbox{r})&=&2C^j_t\bbox{j}_k\ofbboxofr
+ C^{\nabla j}_t\bbox{\nabla}
\times\bbox{s}_k\ofbboxofr,\\
{\mathsf B}_k(\bbox{r})&=&2C^{J0}_tJ_k\ofbboxofr{\mathsfI}
- 2C^{J1}_t\epsilon\cdot\bbox{J}_k\ofbboxofr
+ 2C^{J2}_t\underline{\mathsf J}_k\ofbboxofr
+ C^{\nabla J}_t\epsilon\cdot\bbox{\nabla}\rho_k\ofbboxofr ,\\
M_k(\bbox{r})&=&C^{\tau}_t\rho_k\ofbboxofr, \\
\bbox{C}_k(\bbox{r})&=&C^T_t\bbox{s}_k\ofbboxofr,   \\
\bbox{D}_k(\bbox{r})&=&C^F_t\bbox{s}_k\ofbboxofr,
\enll
for $k=0,1,2,3$.
All coupling constants $C_t$ in Eqs.\ (\ref{potall}) are taken with
$t$=0 for $k=0$ (isoscalars), and with $t$=1 for $k=$1,2,3
(isovectors). Symbol ${\mathsfI}$ is the unit space tensor, and
$\epsilon\cdot\bbox{J}$ stands for the antisymmetric space tensor
with components: $(\epsilon\cdot\bbox{J})_{ab} =
\sum_{c}\epsilon_{acb}\bbox{J}_c$, so that, according to Eq.\
(\ref{tenact}), its action on a vector is obviously the vector
product: $(\epsilon\cdot\bbox{J})\cdot
\hat{\bbox{\sigma}}=\bbox{J}\times \hat{\bbox{\sigma}}$.

The {\pairing} mean-field Hamiltonian has the following
isoscalar and isovector components:
\bnll{hatild}
\breve{h}_0(\bbox{r};s',s)
&=& \breve{\bbox{\Sigma}}_0\ofbboxofr\cdot\hat{\bbox{\sigma}}_{s's}
+   \tfrac{1}{2i}\Big\{\bbox{\nabla}\cdot\breve{\bbox{I}}_0\ofbboxofr\delta_{s's}
+\breve{\bbox{I}}_0\ofbboxofr\delta_{s's}\cdot\bbox{\nabla}\Big\}
-  \bbox{\nabla}\cdot\big[\breve{\bbox{C}}_0\ofbboxofr
\cdot\hat{\bbox{\sigma}}_{s's}\big]\bbox{\nabla}
-  \bbox{\nabla}\cdot \breve{\bbox{D}}_0\ofbboxofr\,\hat{\bbox{\sigma}}_{s's}\cdot \bbox{\nabla},
\label{hatild2} \\
\vec{\breve{h}}(\bbox{r};s',s)
&=& \vec{\breve{U}}(\bbox{r})\delta_{s's}
+   \tfrac{1}{2i}\Big\{\bbox{\nabla}\cdot
\big[\vec{\breve{\mathsf B}}(\bbox{r})\cdot\hat{\bbox{\sigma}}_{s's}\big]
+\big[\vec{\breve{\mathsf B}}(\bbox{r})\cdot\hat{\bbox{\sigma}}_{s's}\big]
\cdot\bbox{\nabla}\Big\}
-  \bbox{\nabla}\cdot\vec{\breve{M}}\ofbboxofr\delta_{s's}\bbox{\nabla}.
\label{hatild1}
\enll
\end{widetext}
Contrary to  the {\particle} Hamiltonian (\ref{ha}), the {\pairing}
Hamiltonian (\ref{hatild}) can be non-hermitian, because potentials
$\breve{\bbox{C}}_0$, $\breve{\bbox{D}}_0$, $\breve{\bbox{I}}_0$,
$\breve{\bbox{\Sigma}}_0$, $\vec{\breve{M}}$, $\vec{\breve{U}}$, and
$\vec{\breve{\mathsf B}}$ are, in general, complex quantities. This
is so, because the {\pairing} density matrix is, in general, not
hermitian. Therefore, the energy functional should be treated as a
functional of both $\hat{\breve{\rho}}$ and $\hat{\breve{\rho}}^+$.

The {\pairing} mean-field Hamiltonian is the functional derivative of
the energy functional over  $\hat{\breve{\rho}}^+$, whereas the
hermitian conjugate Hamiltonian is the functional derivative over
$\hat{\breve{\rho}}$. The {\pairing} densities are, according to
Eqs.\ (\ref{gp1-2}) and (\ref{sp1-2}), functions of
$\hat{\breve{\rho}}$, while the complex conjugate densities are
functions of $\hat{\breve{\rho}}^+$.

When calculating the {\pairing} functional derivatives, one cannot
forget that the {\pairing} density matrix fulfills symmetry condition
(\ref{hc2}), implying that the {\pairing} densities are either
symmetric or antisymmetric functions, Eqs.\ (\ref{symepai}).
Therefore, the calculation  of functional derivatives over either
$\hat{\breve{\rho}}$ or $\hat{\breve{\rho}}^+$ is similar to that
leading to Eqs.\ (\ref{fuder})--(\ref{fuderdiff2}), however,
instead of Eq.\ (\ref{fuderrho}) one has:
\begin{widetext}
\begin{equation}\label{fudersym}
\frac{\delta \hat{\breve{\rho}}^+(\bbox{r}_1s_1t_1,\bbox{r}_2s_2t_2)}
{\delta\hat{\breve{\rho}}^+(\bbox{r}st,\bbox{r}'s't')}
= \delta(\bbox{r}_1-\bbox{r})\delta(\bbox{r}_2-\bbox{r}')\delta_{s_1s}\delta_{s_2s'}\delta_{t_1t}\delta_{t_2t'}
-16ss'tt'\delta(\bbox{r}_1-\bbox{r}')\delta(\bbox{r}_2-\bbox{r})\delta_{s_1-s'}\delta_{s_2-s}\delta_{t_1-t'}\delta_{t_2-t}
\end{equation}
In the expressions for functional derivatives, this gives either
cancellation or addition of terms coming from the two components of
the right-hand side of Eq.\ (\ref{fudersym}). Finally, the
non-vanishing derivatives are
\bnll{fuderbre1}
\frac{\delta \overline{\langle f\vec{\breve{\rho}}^{\ast}\rangle}}
{\delta\hat{\breve{\rho}}^+(\bbox{r}st,\bbox{r}'s't')}&=&
2\delta(\bbox{r}'-\bbox{r}) f(\bbox{r})\delta_{s's}\hat{\vec{\tau}}_{t't}, \\
\frac{\delta\overline{\langle f\breve{\bbox{s}}^{\ast}_0\rangle}}
{\delta\hat{\breve{\rho}}^+(\bbox{r}st,\bbox{r}'s't')}&=&
2\delta(\bbox{r}'-\bbox{r})f(\bbox{r})\hat{\bbox{\sigma}}_{s's}\hat{\tau}^0_{t't},
\enll
\bnll{fuderbre2}
\frac{\delta \overline{\langle f\breve{\bbox{j}}^{\ast}_0\rangle}}
{\delta\hat{\breve{\rho}}^+(\bbox{r}st,\bbox{r}'s't')}&=&
-i\delta(\bbox{r}'-\bbox{r})\big(\bbox{\nabla}f(\bbox{r})
+f(\bbox{r})\bbox{\nabla}\big)
\delta_{s's}\hat{\tau}^0_{t't},
\\
\frac{\delta\overline{\langle f\vec{\breve{\mathsf{J}}}^{\ast}_{ab}\rangle}}
{\delta\hat{\breve{\rho}}^+(\bbox{r}st,\bbox{r}'s't')}&=&
-i\delta(\bbox{r}'-\bbox{r}) \big(\bbox{\nabla}_af(\bbox{r})
+f(\bbox{r})\bbox{\nabla}_a\big)
\hat{\bbox{\sigma}}^b_{s's}\hat{\vec{\tau}}_{t't},
\enll
\bnll{fuderbre3}
\frac{\delta\overline{\langle f\bbox{\nabla}_a\bbox{\nabla}^{\prime}_b
\vec{\breve{\rho}}^{\ast}\rangle}}
{\delta\hat{\breve{\rho}}^+(\bbox{r}st,\bbox{r}'s't')}&=&
-2\delta(\bbox{r}'-\bbox{r}) \bbox{\nabla}_af(\bbox{r})\bbox{\nabla}_b\,
\delta_{s's}\hat{\vec{\tau}}_{t't}, \\
\frac{\delta\overline{\langle f\bbox{\nabla}_a\bbox{\nabla}^{\prime}_b\breve{\bbox{s}}^{\ast}_{0c}\rangle}}
{\delta\hat{\breve{\rho}}^+(\bbox{r}st,\bbox{r}'s't')}&=&
-2\delta(\bbox{r}'-\bbox{r}) \bbox{\nabla}_af(\bbox{r})\bbox{\nabla}_b\,
\hat{\bbox{\sigma}}^c_{s's}\hat{\tau}^0_{t't} ,
\enll
for $a,b,c=x,y,z$.

Using Eqs.\ (\ref{fuderbre1})--(\ref{fuderbre2}) one obtains the
following relations between the potentials defining the {\pairing}
mean-field Hamiltonian (\ref{hatild}) and the local {\pairing}
densities defining the energy density (\ref{chitild}):
\bnll{pairall}
\breve{\bbox{\Sigma}}_0(\bbox{r})&=& 2\breve{C}^s_0\breve{\bbox{s}}_0\ofbboxofr
+ 2\big(\breve{C}^{\Delta s}_0
-\breve{C}^{\nabla s}_0\big)
\Delta\breve{\bbox{s}}_0\ofbboxofr
-  2\breve{C}^{\nabla s}_0\bbox{\nabla}\times
\big(\bbox{\nabla}\times
\breve{\bbox{s}}_0\ofbboxofr\big)
+  \breve{C}^T_0\breve{\bbox{T}}_0\ofbboxofr
+   \breve{C}^F_0\breve{\bbox{F}}_0\ofbboxofr
+\breve{C}^{\nabla j}_0\bbox{\nabla}\times
\breve{\bbox{j}}_0\ofbboxofr, \\
\breve{\bbox{I}}_0(\bbox{r})&=&2\breve{C}^j_0\breve{\bbox{j}}_0\ofbboxofr
+ \breve{C}^{\nabla j}_0\bbox{\nabla}\times
\breve{\bbox{s}}_0\ofbboxofr,\\
\breve{\bbox{C}}_0(\bbox{r})&=& \breve{C}^T_0\breve{\bbox{s}}_0\ofbboxofr,\\
\breve{\bbox{D}}_0(\bbox{r})&=& \breve{C}^F_0\breve{\bbox{s}}_0\ofbboxofr,\\
\vec{\breve{U}}(\bbox{r})&=& 2\breve{C}^{\rho}_1\vec{\breve{\rho}}\ofbboxofr
+  2\breve{C}^{\Delta\rho}_1\Delta\vec{\breve{\rho}}\ofbboxofr
+  \breve{C}^{\tau}_1\vec{\breve{\tau}}\ofbboxofr
+  \breve{C}^{\nabla J}_1\bbox{\nabla}\cdot\vec{\breve{\bbox{J}}}_k\ofbboxofr , \\
\vec{\breve{\mathsf B}}(\bbox{r})&=&2\breve{C}^{J0}_1\vec{\breve{J}}\ofbboxofr{\mathsfI}
-  2\breve{C}^{J1}_1\epsilon\cdot\vec{\breve{\bbox{J}}}\ofbboxofr
+ 2\breve{C}^{J2}_1\underline{\vec{\breve{\mathsf J}}}\ofbboxofr
+  \breve{C}^{\nabla J}_1\epsilon\cdot\bbox{\nabla}\vec{\breve{\rho}}\ofbboxofr ,\\
\vec{\breve{M}}(\bbox{r})&=& \breve{C}^{\tau}_1\vec{\breve{\rho}}\ofbboxofr .
\enll
\end{widetext}
In the case of the zero-range pairing force (\ref{ddpi}), the isovector p-p
potential is proportional to the p-p isovector density while the isoscalar
field  has a very different structure, i.e.,  it
is proportional to the scalar product of spin $\hat{\bbox{\sigma}}$
and the p-p spin density $\breve{\bbox{s}}_0$. This immediately suggests
that there exists a  connection between the isoscalar pairing
and the p-p spin saturation, which is influenced by the spin-orbit splitting.
In this context, let us remind the shell-model study \cite{[Pov98]} which discusses
the relation between the magnitude of the $T$=0 pairing and the
spin-orbit splitting.

\section{The HFB equations}\label{sec7}

Minimization of the energy functional of Eq.\ (\ref{energy}) with
respect to the {\particle} and {\pairing} density matrices, which fulfill
Eqs.\ (\ref{proj})  under auxiliary conditions
\bnll{eq83}
\int {\rmd}^3\bbox{r}\rho_n(\bbox{r})&=& N, \\
\int {\rmd}^3\bbox{r}\rho_p(\bbox{r})&=& Z,
\enll
leads to the HFB equation of the form:
\begin{equation}
\left[\hat{\breve{\mathcal H}},\hat{\breve{\mathcal R}}\right] =0.
\label{HFB}
\end{equation}
The generalized density matrix $\hat{\breve{\mathcal R}}$ is given by Eq.\
(\ref{genden}) and the generalized mean-field Hamiltonian is defined
as
\begin{equation}
\hat{\breve{\mathcal H}} = \hat{\mathcal W}\hat{\mathcal H}\hat{\mathcal W}^+
=\left(\begin{array}{cc}\hat{h}-\hat{\lambda}&\hat{\breve{h}}\\
\hat{\breve{h}}^{+} & -\hat{h}^{TC}+\hat{\lambda}\end{array}\right) ,
\label{genha}
\end{equation}
with the Lagrange multiplier given by
\begin{equation}
\hat{\lambda}=\thalf(\lambda_n+\lambda_p)+\thalf(\lambda_n-\lambda_p)\hat{\tau}^3,
\end{equation}
where $\lambda_n$ and $\lambda_p$ are the neutron and proton Fermi
levels, respectively.

The usual method of solving  the HFB equation (\ref{HFB}) is to solve
in a self-consistent way the eigenvalue problem,
\begin{equation}
\hat{\breve{\mathcal H}}(x',x)\bullet \Phi (x;E) = E\Phi (x';E),
\label{eig}
\end{equation}
for the generalized mean-field Hamiltonian, and then to construct the
generalized density matrix,
\begin{equation}
\hat{\breve{\mathcal R}}(x,x')= \sum_{E\in {\mathcal E}}\Phi(x;E)\Phi^+(x';E),
\end{equation}
as a projection operator onto the set of the quasihole (occupied)
states $\Phi$ belonging to a subset of energy spectrum, $\mathcal E$.
For a local mean-field Hamiltonian, Eq.\ (\ref{eig}) is a system of
eight second-order differential equations, in
general with complex coefficients. Usual four dimensions corresponding
to upper and lower HFB components and to two spin projections are
here multiplied by another factor of two due to the isospin projections.
Altogether, Eq.\ (\ref{eig}) corresponds to a system of
sixteen equations within the domain of real numbers. When specific
symmetry conditions are imposed on solutions, this number can be reduced
in a standard way, see Ref.\ \cite{[Per03a]} for the analysis
pertaining to spherical symmetry.

The energy spectrum of generalized mean-field Hamiltonian has been
discussed in Ref.\ \cite{[Dob84]}. The only difference with the
present case is that here the eigenvalue problems for neutrons and
protons in Eq.\ (\ref{eig}) cannot be separated. It is well known,
that the eigenvalues of $\hat{\breve{\mathcal H}}$ appear in pairs of
opposite signs. For each quasihole state of energy $E$
\begin{equation}
\Phi (\bbox{r}st;E)=\left(\begin{array}{c}\varphi (\bbox{r}st;E)\\
\psi (\bbox{r}st;E)\end{array}\right)
\end{equation}
there exists a quasiparticle state
\begin{equation}
\Phi (\bbox{r}st;-E)=4st\left(\begin{array}{c}\psi^{\ast}(\bbox{r}\,\mbox{$-s-t$};E)\\
\varphi^{\ast}(\bbox{r}\,\mbox{$-s-t$};E)\end{array}\right)
\label{oppos}
\end{equation}
belonging to energy $-E$. In the case of absence of external fields,
bound states (when $\varphi$ and $\psi$ are both localized) exist only
when both Fermi levels, $\lambda_n$ and $\lambda_p$, are negative.
Discrete quasihole energy levels lie within the range ${\mathcal
L}<E<-{\mathcal L}$, where ${\mathcal L}=\max
(\lambda_n,\lambda_p)<0$. The ground-state solution corresponds
to occupying states having negative energies; then the set ${\mathcal
E}$ consists of a number of discrete levels lying inside segment
$({\mathcal L},0)$ and the continuous spectrum with
$-\infty<E<{\mathcal L}$.

Traditionally, one solves Eq.\ (\ref{eig}) for the quasiparticle states
of positive energies rather than for the negative ones. Then, the
discrete spectrum is within the segment $0<E<-{\mathcal L}$ and energies
$E>-{\mathcal L}$ belong to the continuum. Having found the
wavefunctions $\Phi  (\bbox{r}st;E)$ for $E>0$ one uses
Eq.\ (\ref{oppos}) to construct the density matrix, i.e.,
\begin{equation}
\hat{\breve{\mathcal R}}(x,x')= \sum_{E>0}\Phi(x;-E)\Phi^+(x';-E).
\end{equation}
The {\particle} and {\pairing} density matrices are then expressed as
\begin{widetext}
\bnll{eq84}
\hat{\rho}(\bbox{r}st,\bbox{r}'s't')
&=&
16ss'tt'\sum_{E>0}\psi^{\ast}(\bbox{r}\,\mbox{$-s-t$};E)\psi (\bbox{r}'\,\mbox{$-s$}'-t';E), \\
\hat{\breve{\rho}}(\bbox{r}st,\bbox{r}'s't')
&=&
16ss'tt'\sum_{E>0}\psi^{\ast}(\bbox{r}\,\mbox{$-s$}-t;E)\varphi (\bbox{r}'\,\mbox{$-s$}'-t';E).
\enll
\end{widetext}

\section{Conserved symmetries}\label{sec8}

Conserved and broken symmetries are one of the most important
elements of description of many-body systems. Within the mean-field
approach, the theorem about self-consistent symmetries \cite{[RS80]}
tells us that mean-field states may or may not have all the
symmetries of the Hamiltonian, depending on interactions and the
system studied. Within the HFB approach, the symmetry is conserved
when the generalized density matrix $\hat{\mathcal R}$ and the generalized
Hamiltonian $\hat{\mathcal H}$ both commute with the symmetry operator
$\hat{\mathcal U}$, i.e., [$\hat{\mathcal R}$,$\hat{\mathcal U}$]=0 and
[$\hat{\mathcal H}$,$\hat{\mathcal U}$]=0, or
\bnll{consym}
\hat{\mathcal U}\hat{\mathcal R}\hat{\mathcal U}^+ &=& \hat{\mathcal R}, \\
\hat{\mathcal U}\hat{\mathcal H}\hat{\mathcal U}^+ &=& \hat{\mathcal H},
\enll
where
\begin{equation}
\hat{\mathcal U} = \left(\ba{cc} \hat{U} & 0 \\
0      & \hat{U}^* \ea\right) ,
\end{equation}
and $\hat{U}$ is a unitary matrix of the single-particle symmetry operator.
For the ``breve'' representation used in the present study,
the symmetry operator is given) by [cf.\ Eq.\ (\ref{genden})]
\begin{equation}
\hat{\breve{\mathcal U}} = \hat{\mathcal W} \hat{\mathcal U} \hat{\mathcal W}^+
= \left(\ba{cc} \hat{U} & 0 \\
0      & \hat{U}^{TC} \ea\right),
\end{equation}
and then
\bnll{consymbrev}
\hat{\breve{\mathcal U}}\hat{\breve{\mathcal R}}\hat{\breve{\mathcal U}}^+ &=& \hat{\breve{\mathcal R}},
\label{consymbrev-r} \\
\hat{\breve{\mathcal U}}\hat{\breve{\mathcal H}}\hat{\breve{\mathcal U}}^+ &=& \hat{\breve{\mathcal H}}.
\label{consymbrev-h}
\enll
In the previous sections we have presented the most general set
of expressions pertaining to the situation when no symmetries were
{\it a priori} conserved. Below we discuss consequences of conserved
symmetries.

\subsection{Proton-neutron symmetry}\label{sec8a}

The standard case of no proton-neutron mixing can be described
by the conserved proton-neutron symmetry given by
\begin{equation}
\hat{U}_{pn}=i\exp(-i\pi\hat{T}_3)
=i\exp(-\tfrac{i}{2}\pi\hat{\tau}_3)
=\hat{\tau}_3 .
\end{equation}
In other words, the iso-3 signature (multiplied by $i$) is then the
conserved symmetry. Note that conservation of projection of the
isospin on the third axis (the charge conservation) would require
that the iso-3 rotation about an arbitrary angle be conserved, while
the iso-3 signature corresponds only to rotation about $\pi$. Within
the HFB approach, the charge symmetry is broken in the same way as is the
particle number symmetry.

Since the $TC$-transformed symmetry operator reads
$\hat{U}_{pn}^{TC}$=$-\hat{\tau}_3$, we obtain from Eq.\ (\ref{consymbrev-r})
that
\bnll{conservt3}
\hat{\tau}_3 \hat       {\rho}  \hat{\tau}_3 &=&  \phantom{-}\hat       {\rho} , \\
\hat{\tau}_3 \hat{\breve{\rho}} \hat{\tau}_3 &=&            -\hat{\breve{\rho}},
\enll
and analogous properties hold for the mean-field Hamiltonians,
$\hat{h}$ and $\hat{\breve{h}}$, respectively. It is then clear that
without the proton-neutron mixing the {\particle} density matrices
and Hamiltonians have only the $k$=0 and 3 isospin components, while
the {\pairing} ones have (in the ``breve'' representation) only the
$k$=1 and 2 isospin components, cf.\ Eqs.\ (\ref{npden}) and
(\ref{npdenp}).

\subsection{Time-reversal symmetry}\label{sec8b}

In the case of time-reversal invariance, $\hat{\rho}^T$=$\hat{\rho}$
and  $\hat{\breve{\rho}}^T$=$\hat{\breve{\rho}}$, see Eqs.\ (\ref{timerev}),
the {\particle} and {\pairing} densities fulfill additional conditions,
\bnll{symetime}
\label{symetime-a}
{\rho}_0(\bbox{r},\bbox{r}')&=&{\rho}^*_0(\bbox{r},\bbox{r}'),
\quad  \\
\label{symetime-b}
{\rho}_k(\bbox{r},\bbox{r}')&=&-(-1)^{k}{\rho}^*_k(\bbox{r},\bbox{r}'),\\
\label{symetime-c}
{\bbox{s}}_0(\bbox{r},\bbox{r}') &=&
-{\bbox{s}}^*_0(\bbox{r},\bbox{r}'), \quad \\
\label{symetime-d}
{\bbox{s}}_k(\bbox{r},\bbox{r}') &=&
(-1)^{k}{\bbox{s}}^*_k(\bbox{r},\bbox{r}')
\enll
and
\bnll{symetimep}
\label{symetimep-a}
\breve{\rho}_0(\bbox{r},\bbox{r}')&=&\breve{\rho}^*_0(\bbox{r},\bbox{r}'),
\quad  \\
\label{symetimep-b}
\breve{\rho}_k(\bbox{r},\bbox{r}')&=&-(-1)^{k}\breve{\rho}^*_k(\bbox{r},\bbox{r}'),
\\
\label{symetimep-c}
\breve{\bbox{s}}_0(\bbox{r},\bbox{r}') &=&
-\breve{\bbox{s}}^*_0(\bbox{r},\bbox{r}'), \quad \\
\label{symetimep-d}
\breve{\bbox{s}}_k(\bbox{r},\bbox{r}') &=&
(-1)^{k}\breve{\bbox{s}}^*_k(\bbox{r},\bbox{r}'),
\enll
where $k$=1,2,3.
Due to the fact that the $k$=2 Pauli matrix $\hat{\tau}^2$ is imaginary,
the time reversal acts differently on the $k$=2 isovector components
than on the $k$=1,3 components of all isovector densities. At the
first sight, this seems to be a bizarre property. Indeed, the isospin
quantum number is introduced to take into account the fact that there
are two kinds of nucleons in nature, and each kind has its own,
apparently unrelated to one another, time-reversal operation.

However, the use of the standard isospin formalism implies something
more. Namely, the neutron wave function (isospin up) can be obtained
from the proton wave function (isospin down) by an action of the
(real) $\hat{\tau}^1$ Pauli matrix. Therefore, the relative phases of
the neutron and proton wave functions are fixed by the phase
convention that has been used to choose the isospin Pauli matrices.
As a consequence, the time-reversal properties of neutrons and
protons are not any more independent from one another. Of course,
this is not a spurious quirk of the mathematics we use, but a
reflection of a deeper fact that by mixing the neutron and proton
wave functions we introduce complex mixing coefficients that do
affect the time-reversal properties of the mixed wave function.
Conservation of the time reversal means that these mixing
coefficients must follow rules dictated by the time reversal, which
implies differences between the $k$=2 and $k$=1,3 iso-directions.
Therefore, we see here that from basic arguments it follows that
conservation of the time reversal {\em must imply the isospin
symmetry breaking}. The only iso-rotations that are compatible with
the time reversal are those about the $k$=2 iso-axis.
(The influence  of the time-odd fields on the magnitude
of the Wigner energy was pointed out in Ref.~\cite{[Sat98b]}.)

Table \ref{tab1} summarizes properties of {\particle} and {\pairing}
densities under the exchange of their spatial arguments. When no
conserved symmetry is imposed, all densities are complex, and
their real and imaginary parts are either symmetric or antisymmetric.
For conserved time reversal, all densities become either real or
imaginary, and are either symmetric or antisymmetric. Recall that
symmetric parts contribute only to particle, kinetic, spin,
spin-kinetic, and tensor-kinetic local densities, while the antisymmetric
parts contribute only to the current and spin-current local densities.
Therefore, local densities are complex, real, imaginary, or vanishing,
depending on whether time-reversal, proton-neutron, or both symmetries
are conserved. Table \ref{tab2} presents these properties
for all local {\particle} and {\pairing} densities.

In previous studies, e.g., in
Refs.~\cite{[Wol71],[Goo72],[Goo79],[Sat97a],[Ter98]}, the $T$=1
pairing fields were associated with the real part of the pairing
tensor, while the $T$=0 pairing was represented by the imaginary part
of the pairing tensor. Such a structure was obtained for specific
phase conventions and symmetries. On the other hand, as shown in
Table \ref{tab2}, the general case corresponding to no conserved
symmetries (e.g., for rotating states) requires that all the pn
densities be complex.

\begin{table}
\caption[T1]{Symmetries of the {\particle} (left) and {\pairing} (right)
densities in general case (no conserved symmetries imposed),
and in case of the time-reversal symmetry conserved. Real ($\Re$)
and imaginary ($\Im$) parts are symmetric (S) or antisymmetric (A)
under exchange of their spatial arguments,
as indicated in the Table.}\label{tab1}
\begin{ruledtabular}
\begin{tabular}{l|cc|cc@{\quad}||l|cc|cc}
&\multicolumn{2}{c|}{general}&\multicolumn{2}{c@{~}||}{time-rev.}         &
&\multicolumn{2}{c|}{general}&\multicolumn{2}{c}{time-rev.}         \\
density  &$\Re$&\multicolumn{1}{c|}{$\Im$}&$\Re$&$\Im$ &
density  &$\Re$&\multicolumn{1}{c|}{$\Im$}&$\Re$&$\Im$ \\
\hline
$        \rho_0        (\bbox{r},\bbox{r}')$ &  S  &  A  &  S  &  0   &
$\breve{\rho}_0        (\bbox{r},\bbox{r}')$ &  A  &  A  &  A  &  0   \\
$        \rho_2        (\bbox{r},\bbox{r}')$ &  S  &  A  &  0  &  A   &
$\breve{\rho}_2        (\bbox{r},\bbox{r}')$ &  S  &  S  &  0  &  S   \\
$        \rho_{1,3}    (\bbox{r},\bbox{r}')$ &  S  &  A  &  S  &  0   &
$\breve{\rho}_{1,3}    (\bbox{r},\bbox{r}')$ &  S  &  S  &  S  &  0   \\
$        \bbox{s}_0    (\bbox{r},\bbox{r}')$ &  S  &  A  &  0  &  A   &
$\breve{\bbox{s}}_0    (\bbox{r},\bbox{r}')$ &  S  &  S  &  0  &  S   \\
$        \bbox{s}_2    (\bbox{r},\bbox{r}')$ &  S  &  A  &  S  &  0   &
$\breve{\bbox{s}}_2    (\bbox{r},\bbox{r}')$ &  A  &  A  &  A  &  0   \\
$        \bbox{s}_{1,3}(\bbox{r},\bbox{r}')$ &  S  &  A  &  0  &  A   &
$\breve{\bbox{s}}_{1,3}(\bbox{r},\bbox{r}')$ &  A  &  A  &  0  &  A   \\
\end{tabular}
\end{ruledtabular}
\end{table}

\begin{table}
\caption[T1]{Properties of the local {\particle} and {\pairing} densities
in general case (no conserved symmetries imposed), and in case of the
time-reversal, proton-neutron, or both symmetries conserved.
The $k$=0,1,2, or 3 isospin components of densities are complex (C), real
(R), imaginary (I), or zero (0), as indicated in the Table.
}\label{tab2}

\begin{ruledtabular}
\begin{tabular}{l|cccc|cccc|cccc|cccc}
&\multicolumn{4}{c|}{general}    &\multicolumn{4}{c|}{time-rev.}
&\multicolumn{4}{c|}{prot.-neut.}&\multicolumn{4}{c }{both} \\
\hline
k
& 0 & 1 & 2 & 3   & 0 & 1 & 2 & 3   & 0 & 1 & 2 & 3   & 0 & 1 & 2 & 3  \\
\hline
$        \rho_k$
& R & R & R & R   & R & R & 0 & R   & R & 0 & 0 & R   & R & 0 & 0 & R  \\
$        \tau_k$
& R & R & R & R   & R & R & 0 & R   & R & 0 & 0 & R   & R & 0 & 0 & R  \\
$ {\mathsf J}_k$
& R & R & R & R   & R & R & 0 & R   & R & 0 & 0 & R   & R & 0 & 0 & R  \\
$    \bbox{s}_k$
& R & R & R & R   & 0 & 0 & R & 0   & R & 0 & 0 & R   & 0 & 0 & 0 & 0  \\
$    \bbox{T}_k$
& R & R & R & R   & 0 & 0 & R & 0   & R & 0 & 0 & R   & 0 & 0 & 0 & 0  \\
$    \bbox{j}_k$
& R & R & R & R   & 0 & 0 & R & 0   & R & 0 & 0 & R   & 0 & 0 & 0 & 0  \\
$    \bbox{F}_k$
& R & R & R & R   & 0 & 0 & R & 0   & R & 0 & 0 & R   & 0 & 0 & 0 & 0  \\
\hline
$\breve{\rho}_k$
& 0 & C & C & C   & 0 & R & I & R   & 0 & C & C & 0   & 0 & R & I & 0  \\
$\breve{\tau}_k$
& 0 & C & C & C   & 0 & R & I & R   & 0 & C & C & 0   & 0 & R & I & 0  \\
$\breve{{\mathsf J}}_k$
& 0 & C & C & C   & 0 & R & I & R   & 0 & C & C & 0   & 0 & R & I & 0  \\
$\breve{\bbox{s}}_k$
& C & 0 & 0 & 0   & I & 0 & 0 & 0   & 0 & 0 & 0 & 0   & 0 & 0 & 0 & 0  \\
$\breve{\bbox{T}}_k$
& C & 0 & 0 & 0   & I & 0 & 0 & 0   & 0 & 0 & 0 & 0   & 0 & 0 & 0 & 0  \\
$\breve{\bbox{j}}_k$
& C & 0 & 0 & 0   & I & 0 & 0 & 0   & 0 & 0 & 0 & 0   & 0 & 0 & 0 & 0  \\
$\breve{\bbox{F}}_k$
& C & 0 & 0 & 0   & I & 0 & 0 & 0   & 0 & 0 & 0 & 0   & 0 & 0 & 0 & 0  \\
\end{tabular}
\end{ruledtabular}
\end{table}

To summarize this subsection, we now enumerate all non-zero densities
when the time reversal is conserved or not, and/or the proton-neutron
mixing is present or not. By counting as one density each component
of a vector, tensor, or isovector, we obtain the following four
options:

\vspace{1ex}\noindent$1^\circ$ Time reversal broken plus
proton-neutron mixing:
\begin{itemize}
\item[--] 23 real {\particle} isoscalar densities:
$                 \rho_0(\bbox{r})$,
$                 \tau_0(\bbox{r})$,
$          {\mathsf J}_0(\bbox{r})$,
$             \bbox{s}_0(\bbox{r})$,
$             \bbox{T}_0(\bbox{r})$,
$             \bbox{j}_0(\bbox{r})$, and
$             \bbox{F}_0(\bbox{r})$,
\item[--] 69 real {\particle} isovector densities:
$\vec             {\rho}(\bbox{r})$,
$\vec             {\tau}(\bbox{r})$,
$\vec      {{\mathsf J}}(\bbox{r})$,
$\vec         {\bbox{s}}(\bbox{r})$,
$\vec         {\bbox{T}}(\bbox{r})$,
$\vec         {\bbox{j}}(\bbox{r})$, and
$\vec         {\bbox{F}}(\bbox{r})$,
\item[--] 12 complex {\pairing} isoscalar densities:
$     \breve{\bbox{s}}_0(\bbox{r})$,
$     \breve{\bbox{T}}_0(\bbox{r})$,
$     \breve{\bbox{j}}_0(\bbox{r})$, and
$     \breve{\bbox{F}}_0(\bbox{r})$,
\item[--] 33 complex {\pairing} isovector densities:
$\vec{\breve     {\rho}}(\bbox{r})$,
$\vec{\breve     {\tau}}(\bbox{r})$, and
$\vec{\breve{\mathsf J}}(\bbox{r})$,
\end{itemize}

\noindent$2^\circ$ Time reversal conserved plus
proton-neutron mixing:

\begin{itemize}
\item[--] 11 real {\particle} isoscalar densities:
$                 \rho_0(\bbox{r})$,
$                 \tau_0(\bbox{r})$, and
$          {\mathsf J}_0(\bbox{r})$,
\item[--] 30 real {\particle} isovector densities:
$           {\rho}_{1,3}(\bbox{r})$,
$           {\tau}_{1,3}(\bbox{r})$,
$    {{\mathsf J}}_{1,3}(\bbox{r})$,
$       {\bbox{s}}_2    (\bbox{r})$,
$       {\bbox{T}}_2    (\bbox{r})$,
$       {\bbox{j}}_2    (\bbox{r})$, and
$    {{\mathsf J}}_2    (\bbox{r})$,
\item[--] 12 imaginary {\pairing} isoscalar densities:
$     \breve{\bbox{s}}_0(\bbox{r})$,
$     \breve{\bbox{T}}_0(\bbox{r})$,
$     \breve{\bbox{j}}_0(\bbox{r})$, and
$     \breve{\bbox{F}}_0(\bbox{r})$,
\item[--] 33 {\pairing} isovector densities, 22 real:
$\breve     {\rho}_{1,3}(\bbox{r})$,
$\breve     {\tau}_{1,3}(\bbox{r})$,
$\breve{\mathsf J}_{1,3}(\bbox{r})$, and 11 imaginary:
$\breve     {\rho}_2    (\bbox{r})$,
$\breve     {\tau}_2    (\bbox{r})$,
$\breve{\mathsf J}_2    (\bbox{r})$,
\end{itemize}

\noindent$3^\circ$ Time reversal broken, no
proton-neutron mixing:

\begin{itemize}
\item[--] 23 real {\particle} isoscalar densities:
$                 \rho_0(\bbox{r})$,
$                 \tau_0(\bbox{r})$,
$          {\mathsf J}_0(\bbox{r})$,
$             \bbox{s}_0(\bbox{r})$,
$             \bbox{T}_0(\bbox{r})$,
$             \bbox{j}_0(\bbox{r})$, and
$             \bbox{F}_0(\bbox{r})$,
\item[--] 23 real {\particle} isovector densities:
$               {\rho}_3(\bbox{r})$,
$               {\tau}_3(\bbox{r})$,
$        {{\mathsf J}}_3(\bbox{r})$,
$           {\bbox{s}}_3(\bbox{r})$,
$           {\bbox{T}}_3(\bbox{r})$,
$           {\bbox{j}}_3(\bbox{r})$, and
$           {\bbox{F}}_3(\bbox{r})$,
\item[--] 22 complex {\pairing} isovector densities:
$\breve     {\rho}_{1,2}(\bbox{r})$,
$\breve     {\tau}_{1,2}(\bbox{r})$, and
$\breve{\mathsf J}_{1,2}(\bbox{r})$,
\end{itemize}

\noindent$4^\circ$ Time reversal conserved, no
proton-neutron mixing:

\begin{itemize}
\item[--] 11 real {\particle} isoscalar densities:
$               \rho_0(\bbox{r})$,
$               \tau_0(\bbox{r})$, and
$        {\mathsf J}_0(\bbox{r})$,
\item[--] 11 real {\particle} isovector densities:
$             {\rho}_3(\bbox{r})$,
$             {\tau}_3(\bbox{r})$, and
$      {{\mathsf J}}_3(\bbox{r})$,
\item[--] 22 {\pairing} isovector densities, 11 real
${\breve     {\rho}}_1(\bbox{r})$,
${\breve     {\tau}}_1(\bbox{r})$,
${\breve{\mathsf J}}_1(\bbox{r})$, and 11 imaginary
${\breve     {\rho}}_2(\bbox{r})$,
${\breve     {\tau}}_2(\bbox{r})$,
${\breve{\mathsf J}}_2(\bbox{r})$.
\end{itemize}


\section{Conclusions}\label{sec9}

Experimental studies of the heavy $N$$\sim$$Z$ nuclei have
sparked renewed interest  in physics of pn correlations, especially pn pairing.
While the appearance of the $T$=1 pn pairing is
a simple consequence of the charge invariance, in spite of vigorous  research,
no hard evidence  for the   elusive $T$=0 pairing phase has yet been found.
There are conflicting messages coming from calculations based
on the quasiparticle approach.
In some models, the $T$=0 and $T$=1 pairing modes are
mutually exclusive, while
in others they are not. What is clear, however,  that
predictions of
calculations that  impose some
symmetry constraints (which can rule out the presence
of some pairing fields), should be taken with the grain of salt.

In  this work, we propose the most general nuclear energy density
functional which is quadratic in isoscalar and isovector densities.
To this end,
we discuss  the isospin structure of the density matrices and
self-consistent mean fields that appear  in the  coordinate-space  HFB theory
allowing for a  microscopic description of pairing
correlations in all isospin channels. The resulting expressions
incorporate an arbitrary mixing between protons and neutrons. No
particular self-consistent symmetries of the energy density functional
have been imposed, however, the  consequences of the time reversal
and proton-neutron symmetry  are discussed.
The obtained   nuclear energy density functional (\ref{enden}-\ref{chitild})
does not have to be related
to any given local  potential. However,
if the underlying potential is local and velocity-independent, the potential
energy density is invariant with respect to a local gauge transformation.
The resulting  densities appear in certain gauge-invariant  combinations
(\ref{gauge-chiph},\ref{gauge-chitild}) which lead to a significant simplification
of the functional.

The self-consistent wave functions obtained by solving the generalized
HFB equations are not eigenstates of isospin. This is a serious drawback of
the  quasiparticle approach. To cure  this problem, isospin should be restored
by means of, e.g.,   projection techniques.
While this can be carried out in a straightforward manner
for  energy functionals that  are related to a two-body potential,
the restoration of  spontaneously broken
symmetries of a general density functional  poses a  conceptional
dilemma \cite{[Goe93],[Per95],[Fer00],[Yan03]} and a serious challenge that
is left for the future work.

\section{Acknowledgments}
This work has been supported in part by
the Polish Committee for Scientific Research (KBN) under Contract
No.~5~P03B~014~21, by the Foundation for Polish Science (FNP), by
the U.S. Department of Energy under Contract Nos.\
DE-FG02-96ER40963 (University of Tennessee), and DE-AC05-00OR22725
with UT-Battelle, LLC (Oak Ridge National Laboratory).


\end{document}